\documentclass[3p,review]{elsarticle}
\usepackage{lineno,hyperref}

\journal{Composite Structures}

\usepackage{amsmath}
\usepackage{amsthm}
\usepackage{amssymb}
\usepackage{eucal}
\usepackage{mathrsfs}
\usepackage{subfigure}

\usepackage{fix-cm}
\usepackage{bm}
\usepackage{amsfonts}
\usepackage{enumerate}
\usepackage{comment}
\usepackage{amsmath, amssymb}
\usepackage{secdot}
\sectiondot{subsection}
\usepackage{mathrsfs}
\usepackage{mathtools}
\usepackage{type1cm}
\usepackage{pdfpages}
\usepackage{here}
\begin{document}

\begin{frontmatter}

\title{Two-Phase Topology Optimization for Metamaterials with Negative Poisson's Ratio}

\author[mymainaddress]{Daichi Akamatsu}
\author[mymainaddress,mymainaddress2]{Yuki Noguchi }
\author[mymainaddress]{Kei Matsushima}
\author[mymainaddress]{Yuji Sato}
\author[mymainaddress]{Jun Yanagimoto}
\author[mymainaddress,mymainaddress2]{Takayuki Yamada \corref{mycorrespondingauthor}}
\ead{t.yamada@mech.t.u-tokyo.ac.jp}

\cortext[mycorrespondingauthor]{Corresponding author}
\address[mymainaddress]{Department of Mechanical Engineering, Graduate School of Engineering, The University of Tokyo, 2-11-16 Yayoi, Bunkyo-ku, Tokyo 113-8656, Japan.}
\address[mymainaddress2]{Department of Strategic Studies, Institute of Engineering Innovation, Graduate School of Engineering, The University of Tokyo, 2-11-16 Yayoi, Bunkyo-ku, Tokyo 113-8656, Japan.}

\begin{abstract}
Although recent developments in 3D printing technology have made it possible to fabricate metamaterials with characteristic mechanical properties, it is not easy to fabricate complex shapes containing cavities. 
In this study, a composite structure comprising two types of materials without a cavity was optimized. 
Moreover, a mechanical metamaterial with a negative Poisson's ratio that can be fabricated using an additive manufacturing method was developed.
First, a homogenization method that characterizes the properties of composite structures was briefly described. 
Then, an optimization problem to realize a negative Poisson's ratio was formulated, and a level set-based topology optimization method was proposed to solve the abovementioned problem. 
Next, three-dimensional numerical examples are presented to confirm the effectiveness of the proposed method, and the deformation behaviors of the optimized designs are numerically examined. 
Furthermore, a sample containing the optimized design with negative Poisson's ratio for the tensile test was fabricated using a 3D printer. 
{We conducted some experiments to evaluate its mechanical performance.}
\end{abstract}


\end{frontmatter}

\section{Introduction}

{Metamaterials are artificial materials that exhibit unique properties that cannot be found in homogeneous materials.}
Metamaterials have been reported in various fields, including electromagnetic waves \cite{electro} and acoustic waves \cite{sonic}. 
Mechanical metamaterials \cite{mech-meta} that exhibit unique mechanical properties based on their microscopic structures have attracted considerable attention as components that can achieve a macroscopic deformation behavior that is impossible with homogeneous materials. 
A negative Poisson's ratio (PR) is an example of the properties of mechanical metamaterials. 
When an object is loaded in an elastic range, its PR $\nu$ is defined as the ratio of the longitudinal strain to the transverse strain. 
PR is primarily in the range of $0\leq\nu\leq\frac{1}{2}$ for isotropic materials. 
Most metals have a PR of approximately 0.3, whereas natural rubber has a PR of approximately 0.49. 
Generally, PR is positive for almost all materials and material structures. 
Material structures with a negative PR expand in the orthogonal direction when a tensile force is applied, whereas they contract in the orthogonal direction when a compressive force is applied. 
Such structures can be used to improve mechanical properties, improve crack resistance \cite{app1} and fracture toughness \cite{app2}, and tune acoustic response \cite{app3}. 
They have {potential engineering application} in smart antennas, biomedicine, defense, and aerospace \cite{NPR,NPR-appli-example}. 
In \cite{energy}, structures with zero PR were directly designed and layered. 
The experimental results suggested potential applications as internal structures for energy absorption.

Typical mechanical metamaterials comprise microstructures that are periodically arranged. 
The macroscopic properties of metamaterials highly depend on the unit cell structure, which is the smallest unit of the periodic structure.
Therefore, it is important to design the unit cell structure to realize a metamaterial with desired properties.
However, it is not easy to design a metamaterial using a trial-and-error approach because a small change in the design of the unit cell structure can significantly affect the macroscopic properties.

Thus, topology optimization, a flexible structural optimization method, can be used to solve the above mentioned problem. 
Bendsøe and Kikuchi proposed a homogenization-based topology optimization method \cite{kikuchi}.
Since then, various methods have been proposed, including the solid isotropic material with penalization (SIMP) \cite{SIMP} and level set methods \cite{yamada}, and their applications in physics and engineering have been proposed to overcome various challenges, including structural \cite{yamada2}, thermal fluid \cite{ogawa}, and acoustic problems \cite{noguchi}.

In the topology optimization process, it is necessary to analyze the system to be optimized and repeatedly evaluate the objective function, which is the guideline for the design.
Moreover, it is necessary to introduce appropriate evaluation methods because metamaterials are systems comprising multiple periodic arrays of microstructured unit cells.
Therefore, in this study, we introduce a homogenization method that uses the asymptotic expansion of the solution to replace a material structure comprising a periodic structure, such as a metamaterial, with a homogeneous material having equivalent material properties \cite{homo}. 
Macroscopic mechanical properties can be determined using the homogenization method by analyzing the unit cell of a periodic structure.
Darling et al. \cite{sigmund} and Zhang et al. \cite{zhang} used the topology optimization of metamaterials exhibiting negative PRs based on a {homogenization method} to optimally design a two-dimensional (2D) structure comprising a single material and cavity.
Ye et al. \cite{ye} optimized several 2D structures with negative PRs and varying absolute values. 
Vogiatzis et al. \cite{NPR} optimally designed a structure with a negative PR comprising multiple materials and cavity regions in 2D and three-dimensional (3D) problems using a topology optimization based on the {homogenization method}.

A characteristic feature of structures with negative PRs obtained from topology optimization in the previous studies is the mixture of material and cavity regions. 
Material structures with negative PRs are not easy to fabricate using conventional machining because of the periodic arrangement of fine and complex shapes. 
In previous studies \cite{NPR, ye}, material structures with negative PRs obtained from topology optimization were fabricated using an additive manufacturing method, with a higher degree of freedom in the shapes that can be fabricated than the conventional machining. 
{Additive manufacturing methods usually fabricate objects by stacking layered structures divided in equal parts at intervals in a plane parallel to the horizontal plane. }
Thus, if the object has a cavity, it is necessary to fill it with the so-called support material that supports the structure's weight. 
However, in many cases, it is not easy to remove the support material when the object has a complex shape. 
In \cite{Hu}, an optimal structure comprising a single material and cavity with a negative PR in the 2D structural problem was extruded in the thickness direction and fabricated using an additive manufacturing method. 
Similarly, in \cite{NPR}, an optimal structure comprising two materials and a cavity with a negative PR in the 2D structural problem was fabricated using an additive manufacturing method. 
A 3D optimization problem was {also tackled}; however, the 3D optimal structure with material distribution in the thickness direction could not be fabricated because of the difficulty in removing the support material, which is a limitation of their design method.

In this study, we optimize a composite structure of a two-phase material that does not contain a cavity to achieve a negative PR, thus eliminating the requirement for the support material and enabling additive manufacturing. 
Some studies have modeled general structures such as compliant mechanism structures with 3D printers \cite{overhang, compmech}.
However, even if structures comprising microscale structures meet the constraints for modeling with a 3D printer, it is difficult to remove the support material representing the voids as described above, and there are still no papers on this issue to the best of the authors' knowledge.
The cavity-free structure has high rigidity, and there is no heat, noise, or liquid leakage through the cavity. 
Therefore, the cavity-free structure is suitable for application as a component of devices that operating under high loads or fluids. 
Although, several studies\cite{mech-meta, NPR} have been conducted on the optimization of material structures with a negative PR comprising a single material and a cavity or two types of materials and a cavity, there is no study on the optimization of a material structure with a negative PR completely filled with a material without a cavity.
Structures with negative Poisson's ratio are usually made of a solid materials and cavity. 
The negativity is realized by increasing the stiffness ratio of the materials and including cavities.
However, in actual modeling, the inclusion of cavities makes the removal of the support material difficult.
That is, a structure that does not contain cavities is preferred when stacked molding is considered; 
{
	however this is difficult to achieve because the stiffness ratio of the material including cavities is much smaller than cavity-free structures.
}
Therefore, we propose to overcome this problem by introducing topology optimization.

To achieve a negative PR, we formulate a two-step optimization problem. 
In the first step, we optimize a material structure comprising a single material and a cavity region. 
In the second step, we set the obtained optimized structure as the initial design and optimize a material structure comprising two materials without any cavity region. 
{
This two-step algorithm allows us to circumvent local minimal solutions with positive Poisson's ratio.
}
An objective function is formulated in both optimization steps such that the desired homogenized coefficients to realize a negative PR are specified. 
Furthermore, the design sensitivity required for the optimization calculation is derived. 
Thus, we {use} a topology optimization method based on the level set method \cite{yamada} to solve the two-step optimization problem. 
The combination of the level set-based algorithm and homogenization modelling was also used by \cite{NPR}. 
To confirm the effectiveness of the proposed method, we provide an optimized unit cell design without any cavities in a 3D problem and numerically confirm its deformation behavior. 
Furthermore, we fabricated a test piece for a tensile test based on the optimized design using an additive manufacturing method. 
We then performed numerical experiments to demonstrate the deformation behavior of negative PR.

The remainder of this study is organized as follows. 
Section 2 outlines the problem formulation and the proposed homogenization method. 
In Section 3, we formulate the optimization problem of topology optimization and provide an overview of the level set-based topology optimization method \cite{yamada}.
Then, we derive the design sensitivity required to solve the optimization problem.
Section 4 describes the topology optimization algorithm and concrete numerical implementation.
In Section 5, we present numerical examples to validate the proposed method.
Section 6 presents the results of tensile testing to confirm the effectiveness of the proposed method.
Finally, Section 7 presents the conclusion.
\section{Evaluation of mechanical properties of metamaterials based on the homogenization method}

\subsection{Problem setting}\label{sec:2_problem}
\begin{figure}[H]
	\begin{center}
		\includegraphics[scale=0.5]{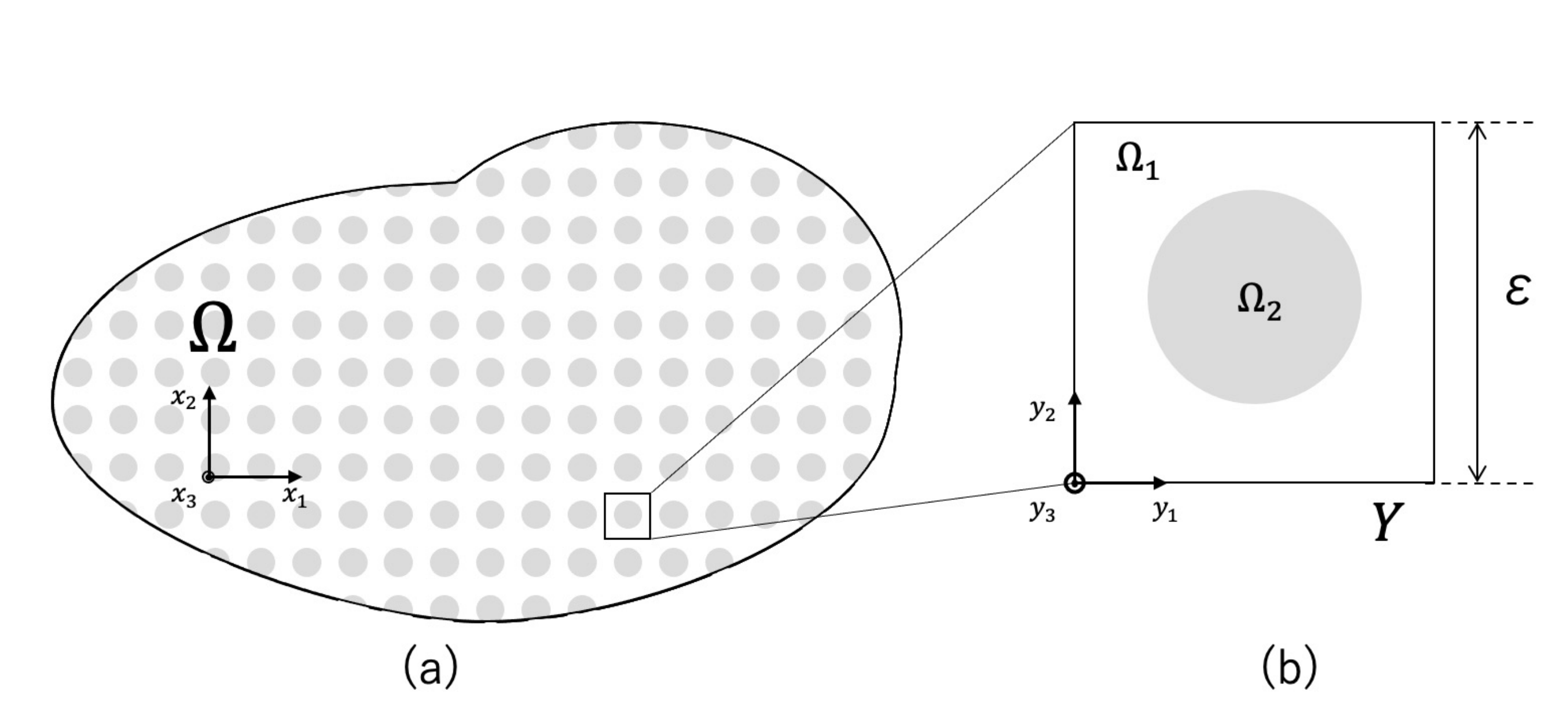}
		\caption{Metamaterial system: (a)the entire system and (b)a unit cell.
	}
		\label{fig:model_problem}
	\end{center}
\end{figure}

As shown in Fig.~\ref{fig:model_problem}, the metamaterial system comprises a region $\Omega$, an array of unit cells. 
In the region $\Omega$, unit cells with a side length of $\varepsilon$ are arranged in a cubic lattice for 3D problems (Fig.~\ref{fig:model_problem}(b)).
Each unit cell is assumed to comprise two isotropic linear elastic materials.
In the unit cell $Y$, the domains $\Omega_1$ and $\Omega_2$ are the regions occupied by materials 1 and 2, respectively.
In Fig.~\ref{fig:model_problem}(b), the white and gray regions are $\Omega_1$ and $\Omega_2$, respectively.

For such a system, the deformation behavior of the metamaterial when subjected to a body force $f(x)$ in the domain $\Omega$ is represented by the displacement field $u_\varepsilon$ as the solution to the following static problem:

\begin{equation}
	\label{gov_macro}
	\left\{ \,
	\begin{aligned}
		-\text{div}\left(C\left(\frac{x}{\varepsilon}\right) e(u_{\varepsilon}) \right) &= f \qquad \text{in} \qquad \Omega, \\
		u_{\varepsilon} &= 0 \qquad \text{on} \qquad \partial \Omega
	\end{aligned}
	\right.
\end{equation} 
where a small deformation of the metamaterial is assumed. 
Here, $e(u)$ is the strain tensor for the vector field $u$, given by $e(u)=\left(\nabla u +(\nabla u)^t\right)/2$; $x$ is the macroscale coordinates for the entire system; and $y=x/\varepsilon$ is the microscale coordinates normalized by the unit cell dimension $\varepsilon$. 
Moreover, $C\left(y\right)$ is a fourth-order elasticity tensor satisfying Hooke's law, which is periodic with respect to the unit cell $\mathit{Y} = (0,1)^N$, i.e., for any $i$-th normal basis vector $e_i$, the coefficients $C(y)$ for $y \in \mathit{Y}$ satisfies the following: 
\begin{equation}
	\label{perio}
	C(y+e_{i}) = C(y), 
\end{equation}
where $N=3$ is the spatial dimension.
The elasticity tensor $C$ can be expressed using Lam\'{e} constants $\lambda$ and $\mu$ as follows:
\begin{align}
	C_{ijkl}=3\lambda\mathscr{J}_{ijkl}+4\mu\mathscr{K}_{ijkl}.
\end{align}
The Lam\'{e} constants $\lambda$ and $\mu$ are expressed using Young's modulus $E$ and PR $\nu$ as follows:
\begin{align}
	\lambda&=\frac{E\nu}{(1+\nu)(1-\nu)} \nonumber \\
	\mu &= \frac{E}{2(1+\nu)}.
\end{align}
Moreover, $\mathscr{J}_{ijkl}$ and $\mathscr{K}_{ijkl}$ are fourth-order tensors defined as follows:
\begin{align}
	\label{param_tensor}
	&\mathscr{I}_{ijkl} \coloneqq \frac{1}{2}(\delta_{ik}\delta_{jl}+\delta_{il}\delta_{jk}), \nonumber\\
	&\mathscr{J}_{ijkl} \coloneqq \frac{1}{N}\delta_{ij}\delta_{kl},\\
	&\mathscr{K}_{ijkl} \coloneqq \mathscr{I}_{ijkl}-\mathscr{J}_{ijkl}, \nonumber
\end{align}
where $\delta_{ij}$ is the Kronecker delta.

{
	We remark that Ali and Simoda \cite{shimoda} have recently proposed a method to reduce computational resources by exploiting the Lam\'{e} constants.
}

Because the unit cell $Y$ comprises two materials, the distribution of $C$ is given as follows:
\begin{equation}
	\label{elastic_tensor}
	C_{\chi}=C_1\chi+C_2(1-\chi),
\end{equation}
where $C_1$ and $C_2$ are the elasticity tensors of materials 1 and 2, respectively; $\chi(y)$ is the characteristic function that is 1 and 0 for materials 1 and 2, respectively, which is explained in detail in the next section.

\subsection{Homogenization method} \label{sec:2_homo}
In this subsection, we introduce the homogenization method \cite{homo} to efficiently evaluate the macroscopic properties of metamaterials comprising microstructures. 
Then, we briefly describe the homogenization method applied to the above described metamaterial system.

If we assume that the dimension of the periodic microstructures $\varepsilon$ is sufficiently small to satisfy $\varepsilon \ll 1$ (Fig.~\ref{fig:model_problem}). 
{In Eq.~(\ref{perio})}, replacing $y$ with $x \text{/} \varepsilon$, we obtain that the map $x \mapsto C(x \text{/} \varepsilon)$ satisfies the periodicity of the period $\varepsilon$ in all coordinate directions $e_1, e_2,~\cdots, e_N$.
We now assume that the solution $u_\varepsilon$ of Eq.~(\ref{gov_macro}) can be asymptotically expanded as follows:
\begin{equation}
	\label{2:homo-1}
	u_\varepsilon (x) = \sum^{+\infty}_{i=0} \varepsilon^i u_i \left(x,\frac{x}{\varepsilon} \right),
\end{equation}
where $u_i(x,y)$ is the asymptotically expanded displacement field, which is a bivariate function of $x$ and $y$ and has periodicity about $y$. 
The spatial gradient of $u_i$ is represented by the gradients $\nabla_x,~ \nabla_y$ for $x$ and $y$ as follows:
\begin{equation}
	\label{2:homo-2}
	\nabla \left( u_i \left( x,\frac{x}{\varepsilon} \right) \right) = (\varepsilon^{-1} \nabla_y u_i + \nabla_x u_i)\left( x,\frac{x}{\varepsilon} \right).
\end{equation}
Substituting Eq.~(\ref{2:homo-2}) into Eq.~(\ref{2:homo-1}), we have
\begin{equation}
	\label{2:homo-3}
	\nabla u_{\varepsilon} (x) = \varepsilon ^{-1} \nabla_y u_0 \left(x,\frac{x}{\varepsilon} \right) + \sum^{\infty}_{i=0} \varepsilon^i (\nabla_y u_{i+1} + \nabla_x u_i ) \left( x,\frac{x}{\varepsilon} \right).
\end{equation}
Substituting Eq.~(\ref{gov_macro}) into Eq.~(\ref{2:homo-3}) and considering the identity for $\varepsilon$ that holds when $\varepsilon \rightarrow 0$, Eq.~(\ref{gov_macro}) can be rewritten as the following problem:
\begin{equation}
	\label{gov_micro}
	\left\{ \,
	\begin{aligned}
		-\text{div}_x\left(C^H(x) e_x(u)\right) &= f \qquad \text{in} \qquad \Omega, \\
		u &= 0\qquad \text{on} \qquad \partial \Omega
	\end{aligned}
	\right.
\end{equation}
where $u(x)=u_0(x)$ represents the main part of $u_\varepsilon$; $\text{div}_x$ and $e_x(u)$ are a partial differential equation and  the strain tensor for the vector field u with respect to $x$, respectively; $C^H$ is a fourth-order tensor called the homogenized coefficient or homogenization elasticity tensor, representing the macroscopic properties. 
$C^H$ is defined as follows:
\begin{align}
	\label{homo_tensor}
	C^H_{ijkl} = \int_{Y} C(e_{ij}+e_y(w_{ij})) : (e_{kl}+e_y(w_{kl}))d {y}, \qquad
	\mathit{i,j,k,l} \in \{1,2,3\},
\end{align}
where $e_{ij}$ is the basis tensor, defined by $e_{ij}=\left( e_i \otimes e_j + e_j \otimes e_i \right)/2$; the second-order tensor $e_y(u)$ is the strain tensor for the vector field $u$  with respect to $y$ as described above; $w_{ij}$ is the displacement field obtained as a response to the unit strain $e_{ij}$ on the unit cell $Y$. 
$w_{ij}$ is obtained by solving the following cell problems defined in $\mathit{Y}$:

\begin{equation}
	\label{gov_weak}
	\left\{ \,
	\begin{aligned}
		\int_{Y} Ce_y(w_{ij} ): e_y(\phi) d {y} + \int_{Y} Ce_{ij} : e_y(\phi)d{y} = 0 \qquad &\text{in} \qquad Y, \\
		y \mapsto w_{ij}({y}) \qquad \qquad \qquad \qquad \qquad &\mathit{Y-periodic},
	\end{aligned}
	\right.
\end{equation}
where $\phi \in \{ H^1_\#(Y)  \}^N$ denotes the test function belonging to the Sobolev space satisfying the periodic boundary conditions.

Using Eqs.~(\ref{homo_tensor}) and (\ref{gov_weak}), we can evaluate the homogenized coefficient from the elasticity tensor $C(y)$ and the displacement field $w_{ij}(y)$ distributed in the unit cell $Y$.

Here, we consider the limit $\varepsilon \rightarrow 0$ in the homogenization method; however, in the experiment, $\varepsilon$ is treated as a finite value as in previous studies \cite{epsilon1, epsilon2}. 
The details are explained in Subsection \ref{sec5:veri}.
\section{Formulation of the optimization problems}

\subsection{Level set-based topology optimization}\label{sec:topo}

The basic idea of topology optimization is to replace a structural optimization problem with a material distribution problem.
A unit cell region $Y$ is defined as a design domain region where the material distributions of materials 1 and 2 are optimized. 
The optimization problem to minimize an objective function $J$ subject to the governing equation expressed in Eq.~(\ref{gov_weak}), where the unit cell structure is set as the design variable, can be formulated as follows:
\begin{align}
	\min_{\chi(y)}~&J \nonumber\\
	\mathrm{subject ~ to}~ \nonumber
	& \mathrm{governing ~ equations ~ in ~}Y \nonumber\\
	& G = \int_{Y} \chi(y) d{y} - V_{\mathrm{max}} \leq 0
	\label{opt_problem}
\end{align}
where the expression of $J$ is specified in the next section; $\chi$ is a characteristic function representing the unit cell structure and is defined as follows:
\begin{equation}
	\chi(y) = 
	\begin{cases}
		1 \qquad \text{if} \qquad \ { y} \,\, \in \Omega_1		 \\
		0 \qquad \text{if} \qquad \ { y} \,\, \in \Omega_2.
	\end{cases}										\label{chi}
\end{equation}
The characteristic function is $\chi=1$ and $0$ in the regions where materials 1 and 2 are present, respectively; $G$ is the volume constraint to limit the domain occupied by material 1, and $V_{\mathrm{max}}$ represents the upper limit of the volume occupied by material 1.

However, the characteristic function can take discontinuous values in an infinitesimal interval. 
The topology optimization problem expressed by Eq.~(\ref{opt_problem}) is known as an ill-posed problem.
In this study, we propose a topology optimization method based on the level set method proposed by Yamada et al. \cite{yamada} and solve the optimization problem of Eq.~(\ref{opt_problem}) after replacing the optimization problem with a well-posed problem via regularization.
The material distribution in the fixed design region $Y$ can be defined using the level set function $\phi({y})$ as follows:
\begin{equation}
	\begin{cases}
		0 < \phi({y}) \leq 1 \qquad &\text{for} \qquad {y} \,\, \in \,\, \Omega_{1}, \\
		\phi({y}) = 0 &\text{for}  \qquad {y} \,\, \in \,\, \partial \Omega, \\
		-1 \leq \phi({y}) < 0 \qquad &\text{for} \qquad {y} \,\, \in \,\, \Omega_{2}.
	\end{cases}
	\label{lev_func}
\end{equation}
In other words, the regions with $\phi >0$ and $\phi<0$ are the regions of materials 1 and 2, respectively, and the isosurface with $\phi=0$ represents their interface $\partial\Omega$.

Using the above material representation by the level set function $\phi$, the optimization problem in Eq.~(\ref{opt_problem}) can be expressed as follows:
\begin{align}
	\min_{\phi} ~&J \nonumber\\
	\mathrm{subject ~ to}~ \nonumber
	& \mathrm{governing ~ equations ~ in ~  }Y, \nonumber\\
	& G = \int_{Y} \chi_\phi({y}) d{y} - V_{\mathrm{max}} \leq 0
	\label{opt_problem2}
\end{align}
where the governing equation is given by Eq.~(\ref{gov_weak}). 
The characteristic function $\chi_\phi$ is defined by $\phi$ as follows:
\begin{equation}
	\chi_{\phi} = 
	\begin{cases}
		1 \qquad \text{for} \qquad \phi( y) \geq 0,\\
		0 \qquad \text{for} \qquad \phi( y) < 0.
	\end{cases}										\label{chi_phi}
\end{equation}
As mentioned in the previous section, using $\chi_\phi$, the distribution of the elasticity tensor $C(\chi_\phi)$ in $Y$ can be expressed as follows:
\begin{equation}
	\label{elastic_tensor}
	C_{\chi_\phi}=C_1\chi_\phi+C_2(1-\chi_\phi),
\end{equation}
where $C_1$ and $C_2$ are the elasticity tensors of materials 1 and 2, respectively.

To solve the optimization problem in Eq.~(\ref{opt_problem2}), we introduce a fictitious time $t$ and replace the process for identifying the optimal solution in the structural optimization problem with a problem of solving the time evolution equation of $\phi$.
In particular, we update the level set function by solving the following reaction--diffusion equation:
\begin{equation}
	\frac{\partial \phi}{\partial t} = -K(J' - \tau \nabla^2_{{y}} \phi) \qquad \text{in} \qquad Y.
	\label{levset}
\end{equation}
where $K (>0)$ is a proportionality constant; $J'$ is the design sensitivity; $\tau (>0)$ is the regularization factor, which can be set to a sufficiently small positive value to properly regularize the optimization problem. 
In this study, periodic boundary conditions are imposed at the external boundary of the unit cell $\partial Y$, and the distribution of the level set function $\phi$ that minimizes the objective function is obtained by solving the reaction--diffusion equation.
%

\subsection{Formulation of the optimization problem}\label{sec:prob2}

In this subsection, we describe the formulation of the optimization problem to achieve negative PRs.

First, we formulate the objective function.
The homogenized coefficient $C^H_{ijkl}$ in the homogenization method described in Subsection \ref{sec:2_homo} is the macroscopic mechanical properties of the metamaterial system. 
The PRs can be expressed as follows:

\begin{equation}
	\nu_{xy} = \frac{C^H_{1122}}{C^H_{1122}+C^H_{2222}}, \,\,\, \nu_{yz} = \frac{C^H_{2233}}{C^H_{2233}+C^H_{3333}}, \,\,	\, \nu_{zx} = \frac{C^H_{3311}}{C^H_{3311}+C^H_{1111}}.
	\label{3d_Poisson}
\end{equation}
Therefore, the objective function $J$ to realize the negative PR is formulated as follows:
\[
J = w_1J_{1} + w_2J_{2} + w_3J_{3} +w_4J_{4} + w_5J_{5} + w_6J_{6}, 
\]
\begin{equation}
	\begin{aligned}
		&J_{1} = (C^H_{1111} - C_{1111}^{t})^2, \,\,\,\,\, J_{2} = (C^H_{2222} - C_{2222}^{t})^2, \,\,\,\,\, J_{3} = (C^H_{3333}-C^t_{3333})^2, \\
		&J_{4} = (C^H_{1122} - C_{1122}^{t})^2, \,\,\,\,\, J_{5} = (C^H_{2233} - C_{2233}^{t})^2, \,\,\,\,\, J_{6} = (C^H_{3311}-C^t_{3311})^2
	\end{aligned}
	\label{3d_obj}
\end{equation}
$J$ comprises $J_{1},~J_{2},~J_{3},~J_{4},~J_{5},$ and $J_{6}$ with prescribed parameters $C^t_{ijkl}$.
Minimizing $J$ means minimizing the difference between the homogenized coefficient $C^H_{ijkl}$ and its target value $C^t_{ijkl}$. 
In this case, by setting the six target values $C^t_{ijkl}$ to satisfy $(C^t_{1122}<0)\,\cap\,$$(C^t_{2233}<0)\,\cap\,$$(C^t_{3311}<0)\,\cap\,$$(C^t_{1122}+C^t_{2222}>0)\,\cap\,$$(C^t_{2233}+C^t_{3333}>0)\cap\,$$(C^t_{3311}+C^t_{1111}>0)$, PR $\nu_{xy},\nu_{yz},\nu_{zx}$ is expected to be negative as per Eq.~(\ref{3d_Poisson}).
Making components such as $C^H_{1122}, C^H_{2233}$, and $C^H_{3311}$ negative means making Poisson's ratio, namely $\nu_{xy}, \nu_{yx}$, and $\nu_{zx}$ negative. 
However, when the objective function was set to simply make these components negative, the optimization problem finds an empty structure with $C^H_{1111}=C^H_{2222}=C^H_{3333}=0$. 
Setting the target values $C^t_{1111}, C^t_{2222}, C^t_{3333} > 0$, we can avoid the trivial local minimal solution.
Therefore, Eq. (\ref{3d_obj}) is set as the objective function in this study.

Using the above objective function $J$ and Eqs.~(\ref{gov_weak}) and (\ref{opt_problem2}), the optimization problem for a 3D structure comprising two materials, materials 1 and 2, is formulated as follows:
\begin{align}
	\min_{\phi} ~&J \nonumber\\
	\mathrm{subject ~ to}~ \nonumber
	& G_{1} = \int_{Y} C(\chi_\phi)e_y({w_{11}} )\cdot e_y(\tilde{{v_{1}}}) d {y} + \int_{Y} C(\chi_\phi)e_{11} \cdot e_y(\tilde{{v_{1}}})d{y} = 0 \nonumber \\
	& G_{2} = \int_{Y} C(\chi_\phi)e_y({w_{22}} )\cdot e_y(\tilde{{v_{2}}}) d {y} + \int_{Y} C(\chi_\phi)e_{22} \cdot e_y(\tilde{{v_{2}}})d{y} = 0 \nonumber \\
	& G_{3} = \int_{Y} C(\chi_\phi)e_y({w_{33}} )\cdot e_y(\tilde{{v_{3}}}) d {y} + \int_{Y} C(\chi_\phi)e_{33} \cdot e_y(\tilde{{v_{3}}})d{y} = 0 \nonumber \\
	& G = \int_{Y} \chi_\phi({y}) d{y} - V_{\mathrm{max}} \leq 0
	\label{opt_problem3d}
\end{align}
The aim of this study is to optimize a cavity-free 3D structure that exhibits a negative PR. 
However, it is not easy to directly solve the optimization problem in Eq.~(\ref{opt_problem3d}) with the cavity-free setting from an initial structure exhibiting a positive PR because it is not easy to sufficiently deform microstructures without cavities.
Therefore, we introduce a two-step optimization problem. 
In the first step, we solve the optimization problem in Eq.~(\ref{opt_problem3d}) for a unit cell structure comprising a solid medium and cavity. 
In other words, we set Young's modulus of two types of elastic materials, $E_1$ and $E_2$, as $E_2 \ll E_1$ based on the Ersatz approach \cite{ersatz}.
Then, in the second step, the optimized design of the first-step optimization is set as an initial configuration of the second-step optimization, and Eq.~(\ref{opt_problem3d}) is solved for a two-phase unit cell structure.
In this step, Young's moduli $E_1$ and $E_2$ are set to represent these two phases.

\subsection{Sensitivity analysis}  \label{sec:sense}

To obtain the optimized design using the optimization method described in Subsection \ref{sec:topo}, 
it is necessary to derive the design sensitivity $J'$ on the right-hand side of Eq.~(\ref{levset}).
In this study, we employed the design sensitivity based on the concept of the topological derivative.
The topological derivative is defined as the rate of change of the objective function when a small inclusion is introduced into a homogeneous base material, and it is defined as follows:
\begin{align}
	\label{derivative}
	DF^{a\rightarrow b} \coloneqq \lim_{\varepsilon \rightarrow +0} \frac{\delta F}{V(\Omega_\varepsilon)},
\end{align}
where  $\Omega_\varepsilon$ is a small spherical inclusion region of radius $\varepsilon$; $\delta F$ is the variation of the objective function, and $V(\Omega_{\varepsilon})$ is the volume of inclusions.

As in the objective function in Eq.~(\ref{3d_obj}), we consider the following objective function $F$ expressed as a function of each component of the homogenized coefficient:
\begin{align}
	\label{obj_sense}
	F = f\left( C^H_{1111}, C^H_{2222},  C^H_{3333},  C^H_{1122},  C^H_{2233},  C^H_{3311} \right).
\end{align}
As per the chain rule of differentiation for each component of the homogenized coefficient,
the topological derivative $D_TF^{a \rightarrow b}$ for the objective function $F$ when an inclusion made of material $b$ is introduced into the domain filled with material $a$ is expressed as follows:
\begin{align}
	\label{dtf}
	D^{a \rightarrow b}_TF &= \frac{\partial f}{\partial C^H_{1111}} D^{a \rightarrow b}_TC^H_{1111} + \frac{\partial f}{\partial C^H_{2222}} D^{a \rightarrow b}_TC^H_{2222} + \frac{\partial f}{\partial C^H_{3333}} D^{a \rightarrow b}_TC^H_{3333} \nonumber \\
	&+ \frac{\partial f}{\partial C^H_{1122}} D^{a \rightarrow b}_TC^H_{1122} + \frac{\partial f}{\partial C^H_{2233}} D^{a \rightarrow b}_TC^H_{2233} + \frac{\partial f}{\partial C^H_{3311}} D^{a \rightarrow b}_TC^H_{3311}
\end{align}
where $D^{a \rightarrow b}_TC^H_{IJKL}$ is the topological derivative for each component of the homogenized coefficient.
It is expressed as follows \cite{topo_deri2}:
\begin{align}
	\label{td_homo}
	D^{a \rightarrow b}_TC^H_{IJKL} =  [e^{IJ}_{ij}+e_{ij}(w^{IJ})]:\mathscr{A}_{ijkl}^{a \rightarrow b}:[e^{KL}_{kl}+e_{kl}(w^{KL})].
\end{align}
$\mathscr{A}_{ijkl}^{a \rightarrow b}$ is the fourth-order tensor called the elastic moment tensor, and it is expressed as follows \cite{Bonnet}: 
\begin{align}
	\label{pala_tensor_3d}
	\mathscr{A}_{ijkl}^{a\to b} \coloneqq \frac{4\pi}{3}\left[ 3\kappa_a\frac{\Lambda_1 -1}{1+\zeta_1(\Lambda_1-1)}\mathscr{J}_{ijkl} + 2\mu_a\frac{\Lambda_2-1}{1+\zeta_2(\Lambda_2-1)}\mathscr{K}_{jikl}\right]
\end{align}
where the subscript $pq$ follows the summation convention. 
Here, the constants $\kappa_m,~\alpha,~\beta,~\Lambda_1,~\Lambda_2,~\Lambda_3,~\zeta_1,~\zeta_2,~\eta_1,~\eta_2,$ and $\eta_3$ are defined as follows:
\begin{align}
	\label{param}
	&\kappa_m=\frac{E_m}{3(1-2_nu_m)}, \,\alpha=\frac{1+\nu_a}{1-\nu_a}, \,\beta=\frac{3-\nu_a}{1+\nu_a}, \nonumber \\
	&\Lambda_1=\frac{\kappa_b}{\kappa_a}, \,\Lambda_2=\frac{\mu_b}{\mu_a}, \,\Lambda_3=\frac{E_b}{E_a},\,\zeta_1 = \frac{1+\mu_a}{3(1-\mu_a)},\, \zeta_2=\frac{8-10\nu_a}{15(1-\nu_a)},  \\
	&\eta_1 = \frac{1+\nu_b}{1+\nu_a}, \,\eta_2 = \frac{1-\nu_b}{1-\nu_a}, \,\eta_3 = \frac{\nu_b(3\nu_a-4)+1}{\nu_a(3\nu_a-4)+1}\nonumber
\end{align}
where $E_m,~\nu_m,$ and $\mu_m$ are Young's modulus, PR, and shear modulus of material $m(=a,b)$, respectively; $\mathscr{I}_{ijkl},~\mathscr{J}_{ijkl}$, and $\mathscr{K}_{ijkl}$ are fourth-order tensors defined in Eq.~(\ref{param_tensor}).

%
\section{Numerical implementation}

In this section, we describe a numerical implementation based on the formulation of the optimization problem in the previous section.

\subsection{Optimization algorithm}

As mentioned in Subsection 3.2, we use a two-step optimization problem to determine the target cavity-free structure.

{
	Although the original level set-based algorithm \cite{yamada} should suffice to realize negative PRs, we observe that the naive algorithm typically falls into solutions (structures) with positive Poisson's ratio. This is possibly because the objective function $J$ is the summation of the multiple functions $J_1,\ldots,J_6$, i.e., this is a multi-objective optimization problem. This means that it is more likely that the algorithm yields local optimal solutions. Although the desired global solution $J=0$ should be attained when appropriate coefficients $w_1,\ldots,w_6$ are given, it is quite difficult to tune the parameters by trial and error. Instead, we introduce an auxiliary design variable (i.e., cavity) at the first step. This allows us to realize negative PRs more easily because of the high-contrast material constants. At the second step, we remove the cavity from the structure and continue the optimization. 
}

In the first step, where the unit cell structure with cavities is optimized, the optimization calculation is based on the following algorithm.
\begin{enumerate}[Step 1.]
	\item: Set the initial value of the level set function and initialize each variable.
	\item: Solve the governing equations given by Eq.~(\ref{gov_weak}) using the finite element method (FEM) to obtain the state fields $w_{11}, w_{22},$ and $w_{33}$.
	\item: Calculate the homogenized coefficients $C^H_{1111}, C^H_{2222}, C^H_{3333}, C^H_{1122}, C^H_{2233},$ and $C^H_{3311}$ from the state fields using Eq.~(\ref{homo_tensor}).
	\item: Compute the objective function given by Eq.~(\ref{3d_obj}) and determine the convergence.
	\item: Calculate the design sensitivity given by Eq.~(\ref{dtf}).
	\item: Update the level set function by the reaction--diffusion equation given by Eq.~(\ref{levset}) using the FEM.
	\item: Return to Step 2.
\end{enumerate}

In the second-step optimization, where the unit cell structure without cavities is optimized, the optimization calculations are primarily the same as in the first-step algorithm. 
However, the initial structure in Step 1 is used as the optimal structure obtained in the first-step optimization, and the target values of each homogenization factor and volume fraction in the objective function in Step 4 are changed from the first step.

In the two-step problem, we performed the design model creation, finite element analysis, and sensitivity analysis using FreeFem++ \cite{freefem++}, an open-source FEM solver. 
The numerical example presented in Subsection \ref{sec5:veri} was confirmed using FEM calculations.

%
\subsection{Approximating the characteristic function}
Rather than using the characteristic function $\chi_\phi$ defined in Eq.~(\ref{chi_phi}), we used an approximated Heaviside function $H(\phi)$ for numerical calculation: 
\begin{equation}
	H(\phi)=
	\begin{cases}
		0&(\phi<-d)\\
		\frac{1}{2}+\frac{\phi}{d}\Bigl\{ \frac{15}{16} -\frac{\phi^2}{d^2} \bigl\{ \frac{5}{8} - \frac{3}{16} \frac{\phi^2}{d^2} \bigr\} \Bigr\}&(-d\leq \phi < d)\\
		1&(d\leq\phi)
	\end{cases}
	\label{eq:heviside}
\end{equation}
where $w$ is the transition width of the approximated Heaviside function, which should be sufficient.

%
\section{Numerical examples}\label{5num}

In this section, we evaluate the validity and effectiveness of the proposed method by presenting multiple numerical examples, including optimized designs exhibiting negative PRs. 
First, we offer an optimized design of the unit cell structure comprising a solid medium and cavity regions, obtained from the first-step optimization. 
Then, we present an optimized design comprising two materials without any cavity region obtained from the second-step optimization, and its deformation properties are demonstrated. 

Each of the two-phase materials that make up the unit cell is set to a resin material that can be used in a 3D printer.
The two materials are selected to be a relatively stiff material and a material sufficiently less stiff than the material for the 3D printer because it is known from Young's modulus settings used in previous studies that the greater the ratio of stiffness of two materials, the greater the PR.

Next, as material settings, material 1 is a resin material with high stiffness, whereas material 2 is a resin material with low stiffness and rubber-like properties.
The Young's modulus and PR of these resin materials were set to $E=2,000~\text{MPa}$ and $\nu=0.3$ \cite{stratasys} for material 1 and $E=1.0~\text{MPa}$ and $\nu=0.45$ \cite{tango} for material 2, corresponding to the materials VeroClear and TangoPlus provided by Stratasys, respectively.

\subsection{Results from the first-step optimization}\label{sec5:opt_3d}

First, we demonstrate the optimization results for the unit cell structure comprising a single material and a cavity in a 3D problem.
As explained in Subsection \ref{sec:prob2}, a cavity region is allowed in this optimization; material 2 is set as a material with low stiffness to allow the cavity. 
In the optimization calculations shown below, we used the nondimensionalized Young's modulus, with Young's moduli of materials 1 and 2 set to $E_1=2.0$ and $E_2=1.0~\times~10^{-3}$, respectively. 
The Young's modulus corresponding to the cavity is set to $E_2=2.0~\times~10^{-5}$.
The PR was then set to $0.3$ for material 1 and cavity.
The regularization factor for the reaction--diffusion equation was then set to $\tau=1~\times~ 10^{-4}$.
As a unit cell, we set a cube with a side length of 1 in the $y$-coordinate. 
A cube comprising material 1 with a spherical hole of radius 0.25 at the center of the unit cell is set as an initial structure of the optimization.

The homogenized coefficients of the initial structure are $(C^H_{1111}, ~C^H_{2222},~C^H_{3333}) = (2.32,~2.32,~2.32),~(C^H_{1122},~C^H_{2233},\\
~C^H_{3311}) = (0.950,~0.950,~0.950)$, and the macroscopic PRs are $(\nu_{xy},~\nu_{yz},~\nu_{zx}) = (0.290,~0.290,~0.290)$. 
The volume ratio in the initial structure is $0.934$.
For this initial structure, the target values of the homogenized coefficient $C^t_{ijkl}$ were set to $C^t_{1111}=C^t_{2222}=C^t_{3333}=0.18,~ C^t_{1122}=C^t_{2233}=C^t_{3311}=-0.045$, respectively, and the objective function $J$ was minimized. 
The target values for each homogenization factor were determined by the trial-and-error method while trying various objective functions.
The target values for the homogenization factor in the Subsection 5.2 were set in the same manner.
The volume constraint was introduced such that the volume ratio would be $V_\mathrm{max}=0.60$, and the weighting factors of the objective function were set to $w_1=1.0,~w_2=1.0,~w_3=1.0,~w_4=2.0, 
~w_5=2.0, $ and $w_6=2.0$.

\begin{figure}[H]
	\begin{center}
		\includegraphics[scale=0.5]{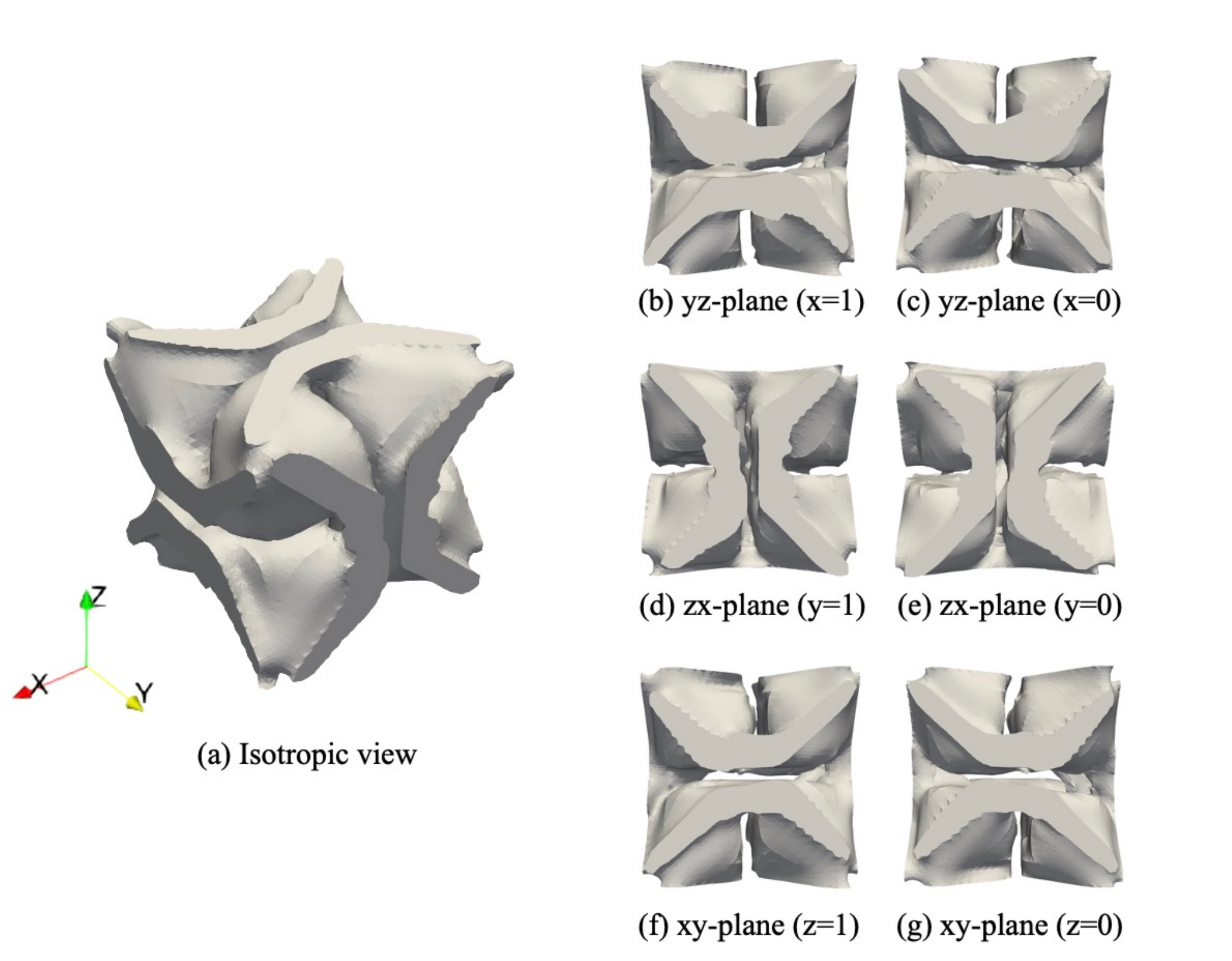}
		\caption{Optimal configuration after the first-step optimization. }
		\label{fig:3D-opt}
	\end{center}
\end{figure}

\begin{figure}[H]
	\begin{center}
		\includegraphics[scale=0.5]{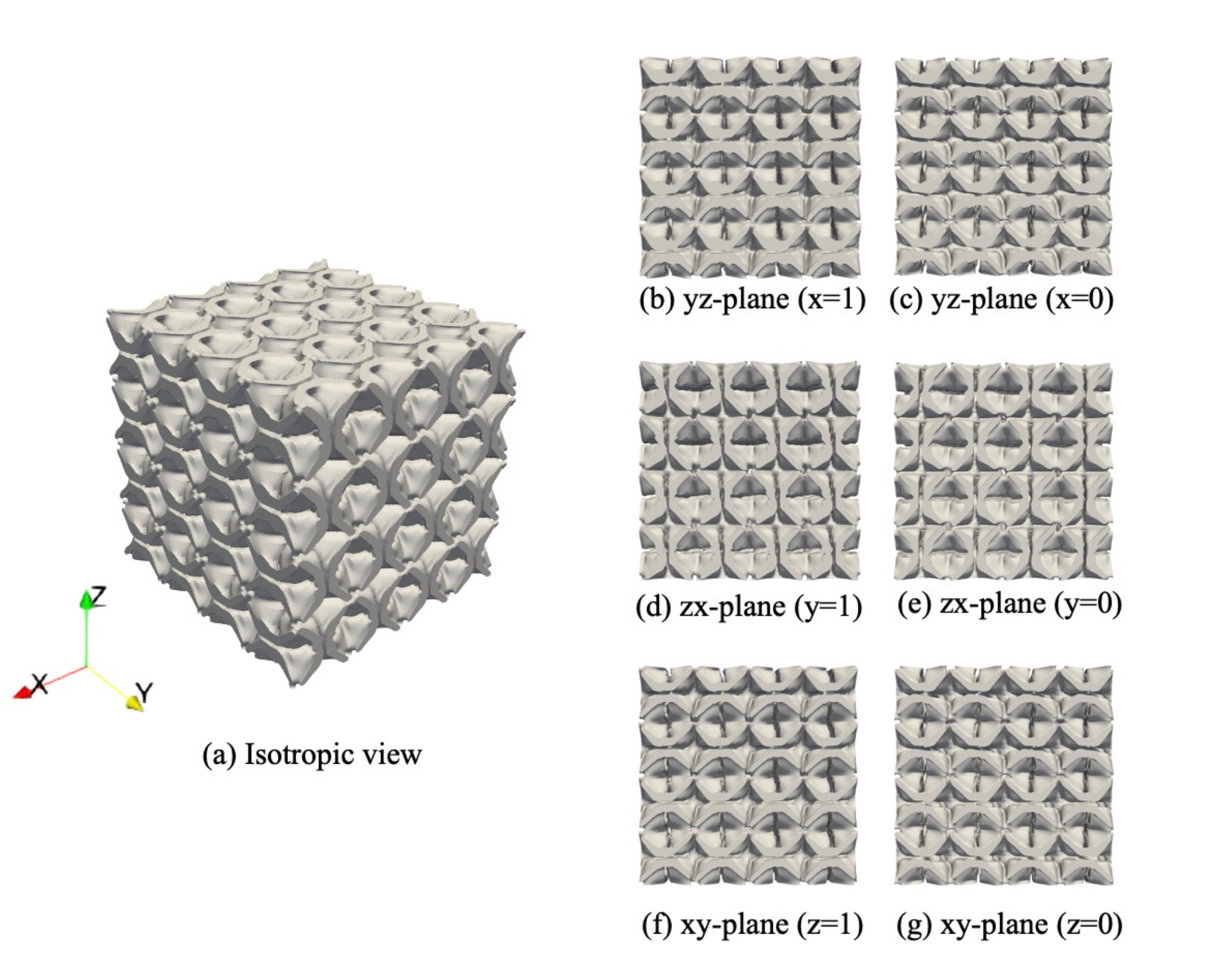}
		\caption{4$\times$4$\times$4 optimal configuration to satisfy the periodic boundary condition.}
		\label{fig:3D-opt-64}
	\end{center}
\end{figure}

Fig.~\ref{fig:3D-opt}(a) shows the optimal structure obtained from this optimization, and Figs.~\ref{fig:3D-opt}(b)-(g) show each surface.
The homogenized coefficient of the optimal structure is 
$(C^H_{1111},~C^H_{2222},~C^H_{3333}) = (0.188,~0.184,~0.149), 
\\(C^H_{1122},~C^H_{2233},~C^H_{3311}) = (-0.0485,~-0.0428,~-0.0418)$, and the macroscopic PR is $(\nu_{xy}, ~\nu_{yz},~\nu_{zx}) = (-0.358,~-0.403,~-0.286)$.
It can be seen that each homogenization factor $C^H_{ijkl}$ is approaching the target value $C^t_{ijkl}$. Especially for $C^H_{3333}$, the error from the target value is a little large, but this is thought to be due to the fact that it oscillates around the optimal solution.
The volume ratio is $0.628$, indicating that the volume constraint is approximately satisfied.
Fig.~\ref{fig:3D-opt-64} shows the design of a metamaterial comprising the $4\times4\times4$ unit cells of the optimized design. 
Fig.~\ref{fig:3D-opt-64}(a) shows the isotropic view, and  Figs.~\ref{fig:3D-opt-64}(b)-(g) show each surface structure.
As shown in Figs.~\ref{fig:3D-opt} and \ref{fig:3D-opt-64}(a), multiple unit cell structures containing cavities are arranged; thus, support material must be used for additive manufacturing.
However, the support material is not easy to remove because of the structure complexity, thus making additive manufacturing with a 3D printer difficult.

\subsection{Results at the second-step optimization}\label{sec5:opt_3d_novoid}

Next, we present the optimization results for the unit cell structure comprising two materials without cavities in a 3D problem.

Similar to the previous section, we used the nondimensionalized Young's modulus in this optimization. 
Young's moduli of materials 1 and 2 are set to $E_1=2.0$ and $E_2=1.0~\times~10^{-3}$, respectively, corresponding to VeroClear and TangoPlus, respectively.

Similar to the initial structure, we use the optimal structure of Fig.~\ref{fig:3D-opt}(a) obtained in Subsection \ref{sec5:opt_3d}.
Material 1 is placed in the region where the material is placed in Fig.~\ref{fig:3D-opt}(a), and material 2 is placed in the hollow region.
The homogenized coefficients of the initial structure are $(C^H_{1111},~C^H_{2222},~C^H_{3333}) = (0.238,~0.228,~0.199),~(C^H_{1122},~C^H_{2233},~C^H_{3311}) = (-0.0291,~-0.0256,~-0.0232)$, and the macroscopic PRs are $(\nu_{xy},~\nu_{yz},~\nu_{zx}) = (-0.146,~-0.147,~-0.109)$. 
The placement of TangoPlus in the cavity region of the optimal structure obtained in the first-step optimization results in negative PRs; however, the absolute value of the corresponding PRs is smaller.
The volume ratio in the initial structure is $0.628$.
For this initial structure, the target values of the homogenized coefficient $C^t_{ijkl}$ were set to $C^t_{1111}=C^t_{2222}=C^t_{3333}=0.20,~C^t_{1122}=C^t_{2233}=C^t_{3311}=-0.06$. 
For the volume ratio, we introduced a volume constraint such that the upper limit was $V_\mathrm{max}=0.45$, and the weight coefficients of the objective function were set to $w_1=1.0,~w_2=1.0,~w_3=1.0,~w_4=2.0,~w_5=2.0,$ and $w_6=2.0$. 
Consequently, the objective function $J$ was minimized under these conditions.

\begin{figure}[H]
	\begin{center}
		\includegraphics[scale=0.5]{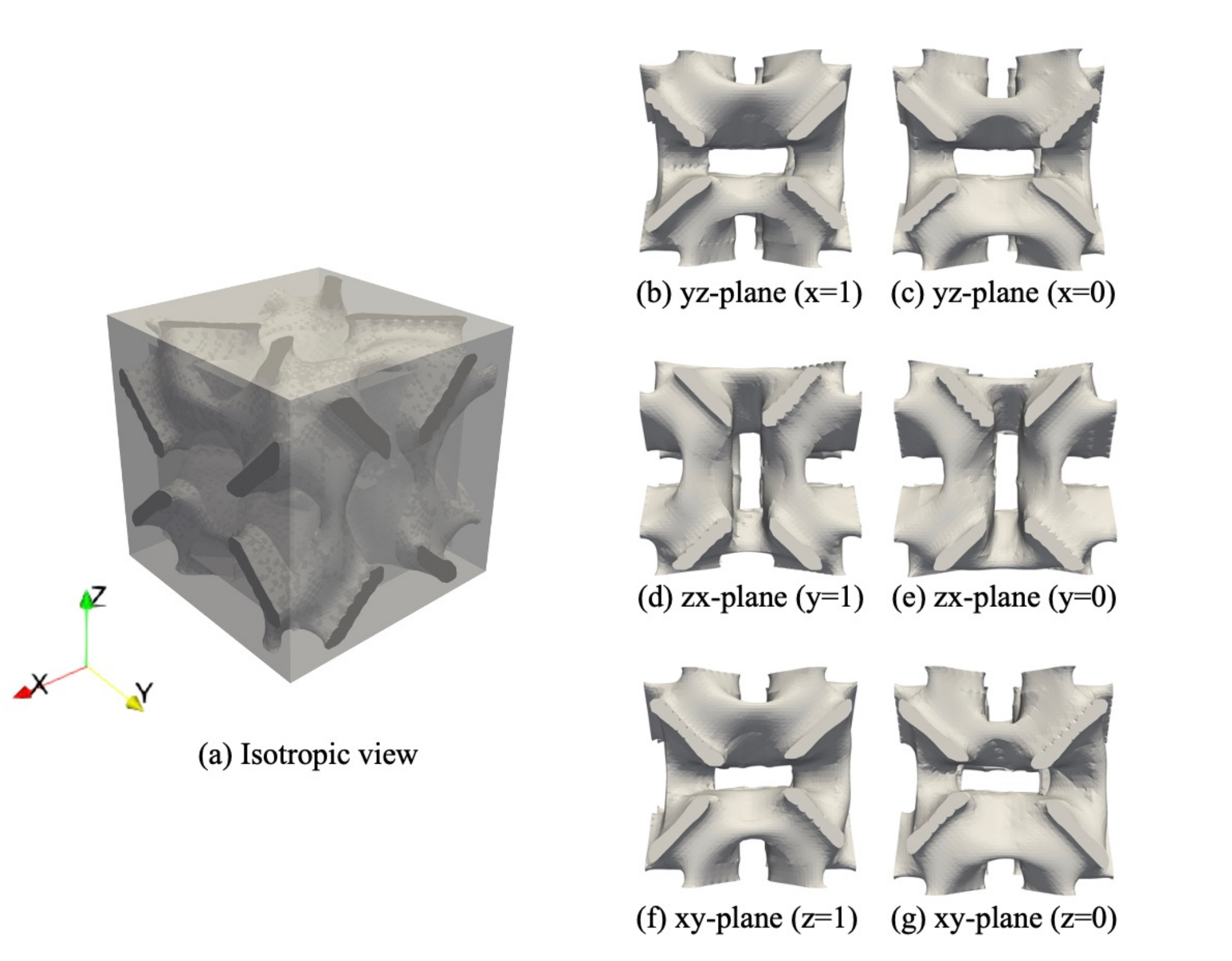}
		\caption{Optimal configuration after the second-step optimization. In (b)-(g), only material 1 is shown for visibility.}
		\label{fig:3D-novoid-opt}
	\end{center}
\end{figure}
\begin{figure}[H]
	\begin{center}
		\includegraphics[scale=0.5]{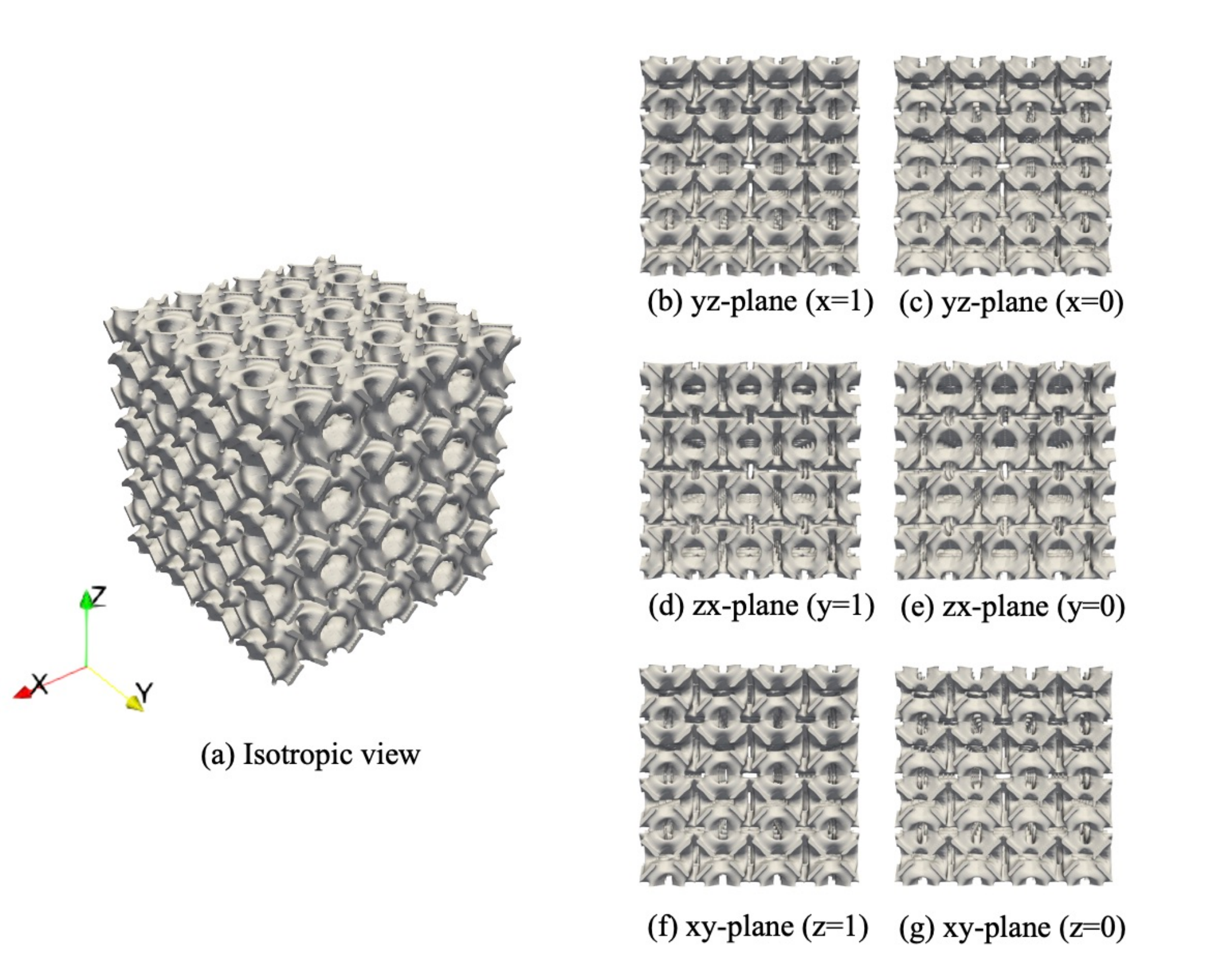}
		\caption{4$\times$4$\times$4 optimal configurations satisfying the periodic boundary condition. Only material 1 is shown for visibility.}
		\label{fig:3D-novoid-opt-64}
	\end{center}
\end{figure}

Fig.~\ref{fig:3D-novoid-opt}(a) shows the optimal structure. 
The gray and translucent  domains represent the object regions occupied by materials 1 and 2, respectively. 
Figs.~\ref{fig:3D-novoid-opt}(b)-(g) show each surface structure; only the structure of material 1 is shown for visibility.
The homogenized coefficients of the optimal structure are
$(C^H_{1111},~C^H_{2222},~C^H_{3333}) = (0.200,~0.200,~0.200),~(C^H_{1122},~C^H_{2233},~C^H_{3311}) = (-0.0595,~-0.0592,~-0.0579)$, and the macroscopic PRs are $(\nu_{xy},~\nu_{yz},~\nu_{zx}) = (-0.423,~-0.422,~-0.409)$.
In all cases, the error from the target value $C^t_{ijkl}$ is less than $5\%$, and PRs are negative and have larger absolute values than the initial structure.
Moreover, the volume ratio is $0.449$, indicating that the volume constraint is satisfied.
Fig.~\ref{fig:3D-novoid-opt-64}(a) shows the geometry of the optimal structure of the unit cell arranged in $4~\times~4~\times~4$ to satisfy the periodic boundary condition. 
Figs.~\ref{fig:3D-novoid-opt-64}(b)-(g) show each surface structure; only material 1 is observed for visibility.
The optimal structure is characterized by the fact that material 1 is connected as a framework.
Moreover, a negative PR can be achieved with a structure that does not contain cavities, as in the objective.

\subsection{ Validation of the optimization calculation \label{sec5:veri}}
\subsubsection{Macroscopic deformation behavior with the results in Subsection \ref{sec5:opt_3d}}

\begin{figure}[H]
	\begin{tabular}{cc}
		\begin{minipage}[t]{0.35\hsize}
			\centering
			\includegraphics[scale=0.5]{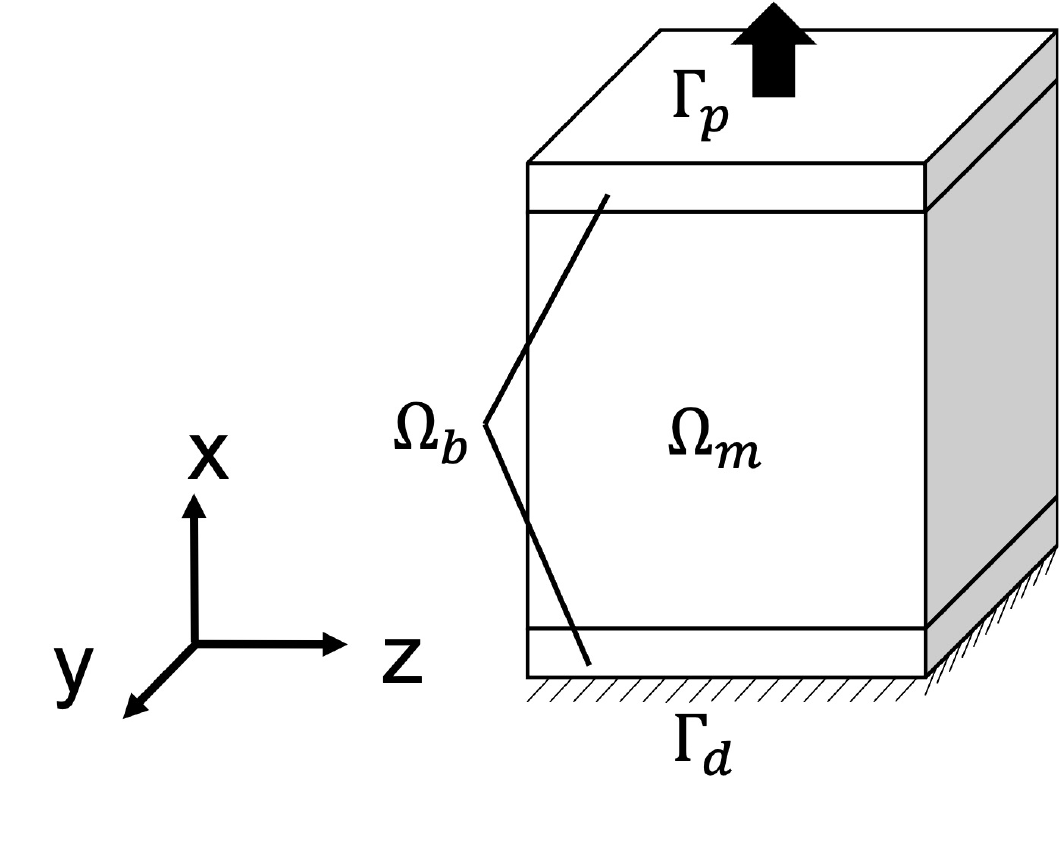}
			\caption{Computational model for analyzing the deformation behavior of negative PR.}
			\label{fig:model-velidation}
		\end{minipage}&
		\begin{minipage}[t]{0.6\hsize}
			\centering
			\includegraphics[scale=0.5]{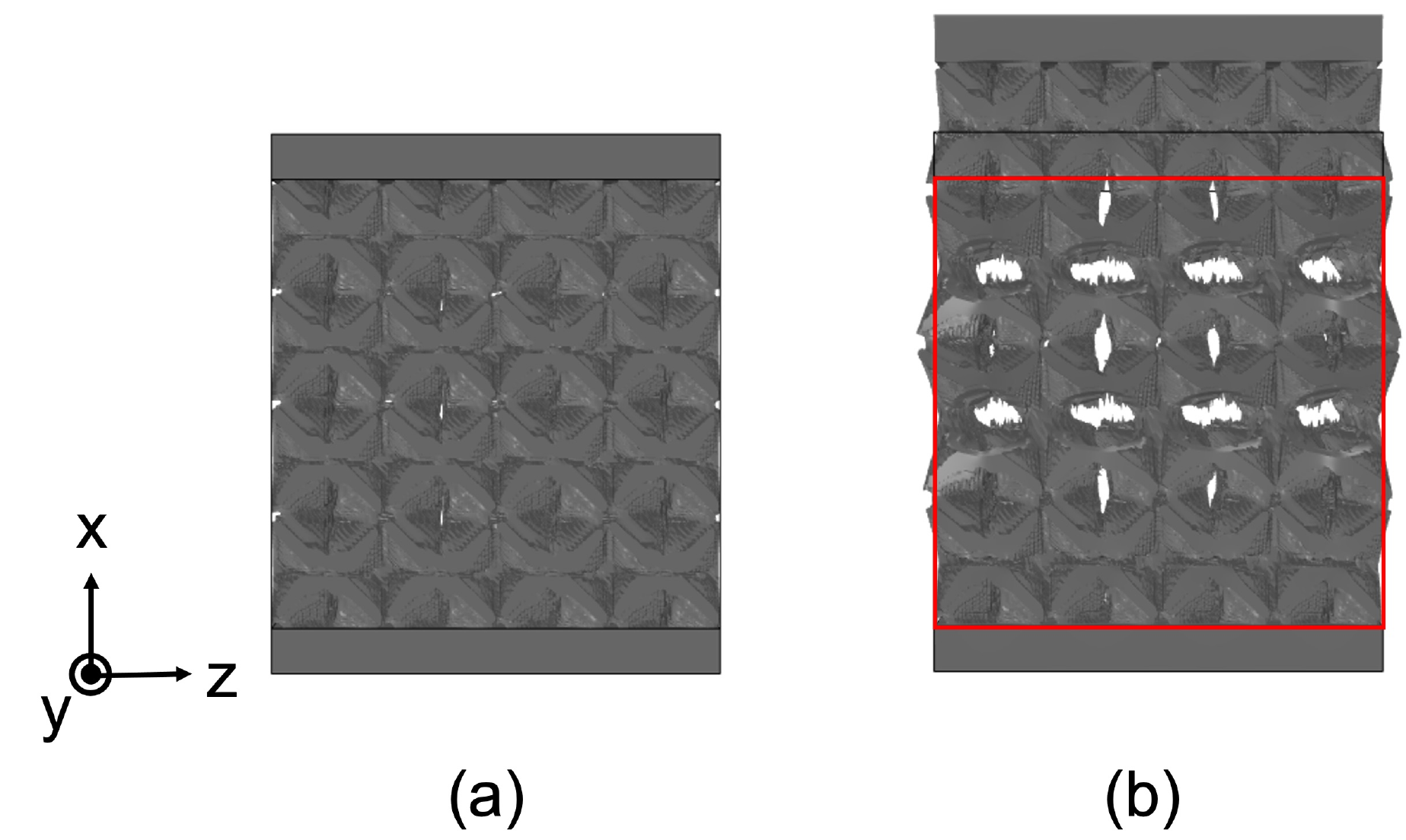}
			\caption{Validation of a 3D metamaterial with negative PRs containing cavity region: (a) shape before deformation and (b) shape after deformation.}
			\label{fig:3D-comsol}
		\end{minipage}
	\end{tabular}
\end{figure}

In this subsection, we confirm the optimal structure obtained in the Subsection \ref{sec5:opt_3d}.
We developed a metamaterial by periodically arranging multiple optimal structures and examined its deformation behavior by applying a prescribed displacement to the metamaterial using FEM calculations.
Fig.~(\ref{fig:model-velidation}) shows the model for confirming the optimal structure.  
The domain $\Omega_b$ comprising bulk material was set above and below the domain $\Omega_m$ comprising multiple unit cell structures.
The boundary conditions were then applied by setting $\Gamma_d$ and $\Gamma_p$ as fixed and prescribed displacement boundaries, respectively, where the displacement was fixed at $(1.0,~0,~0)$. 
The unit cell size was set to $10$ mm, and $4~\times~4~\times~4$ unit cells of the optimized designs were periodically arranged in a cubic lattice.
The domain $\Omega_b$ is the bulk material of material 1 with a dimensions $3~\times~10~\times~10~\mathrm{mm^3}$ and the behavior of the metamaterials was confirmed when subjected to strain along the $x$-axis.

Fig.~\ref{fig:3D-comsol}(a) shows the optimal structure of the unit cell shown in Fig.~\ref{fig:3D-opt}(a) as in Section \ref{fig:3D-opt-64}, arranged in $4~\times~4~\times~4$ to satisfy the periodic boundary condition. 
The black and white domains are material 1 and hollow regions, respectively.
Fig.~\ref{fig:3D-comsol}(b) shows the behavior of the structure shown in Fig.~\ref{fig:3D-comsol}(a) when subjected to tensile boundary displacement along the $x$-axis. 
The red box shows the structural region before deformation. 
When strain is applied in the tensile direction along the $x$-axis, the material comprising the unit cell, i.e., the black region in Fig.~\ref{fig:3D-comsol}, undergoes strain in the orthogonal direction because of the opening along the $z$-axis. 
Moreover, the entire structure consistently expands with the PR value $\nu_{zx}=-0.286$. 
As mentioned in the Subsection \ref{sec5:opt_3d}, the PRs of the optimal structure are $(\nu_{xy}, ~\nu_{yz},~\nu_{zx}) = (-0.358,~-0.403,~-0.286)$, which are almost the same for all three directions, indicating that negative PRs are indeed realized.
In Fig.~\ref{fig:model-velidation}, Poisson's ratio $\nu_{zx}$ was calculated by calculating the transverse strain from the displacements of the left and right planes when vertical strain was applied in the x direction, and Poisson's ratio $\nu_{xy}$ was calculated from the displacements of the front and back planes.
The Poisson's ratio $\nu_{yz}$ was calculated by performing the same calculation for the shape when tensile displacement was applied in the y-direction.
The result is $( \nu_{xy}, \nu_{yz}, \nu_{zx})=(-0.15,-0.22,-0.20)$, indicating that a negative Poisson's ratio is indeed realized. 
In the homogenization theory, the structure is assumed to be composed of an infinite number of unit cells.
However, we arrayed finite numbers ($4 \times 4 \times 4$) of unit cells due to the limitation of the printing arruracy of the 3D printer. 
Therefore, the absolute value of Poisson's ratio is considered to be smaller than the aforementioned value.

\subsubsection{Macroscopic deformation behavior with the results in Subsection \ref{sec5:opt_3d_novoid}}

In this subsection, we confirm the optimal structure obtained in Subsection \ref{sec5:opt_3d_novoid} by examining the representative behavior when strain is applied along the $x$-axis.

\begin{figure}[H]
	\begin{center}
		\includegraphics[scale=0.5]{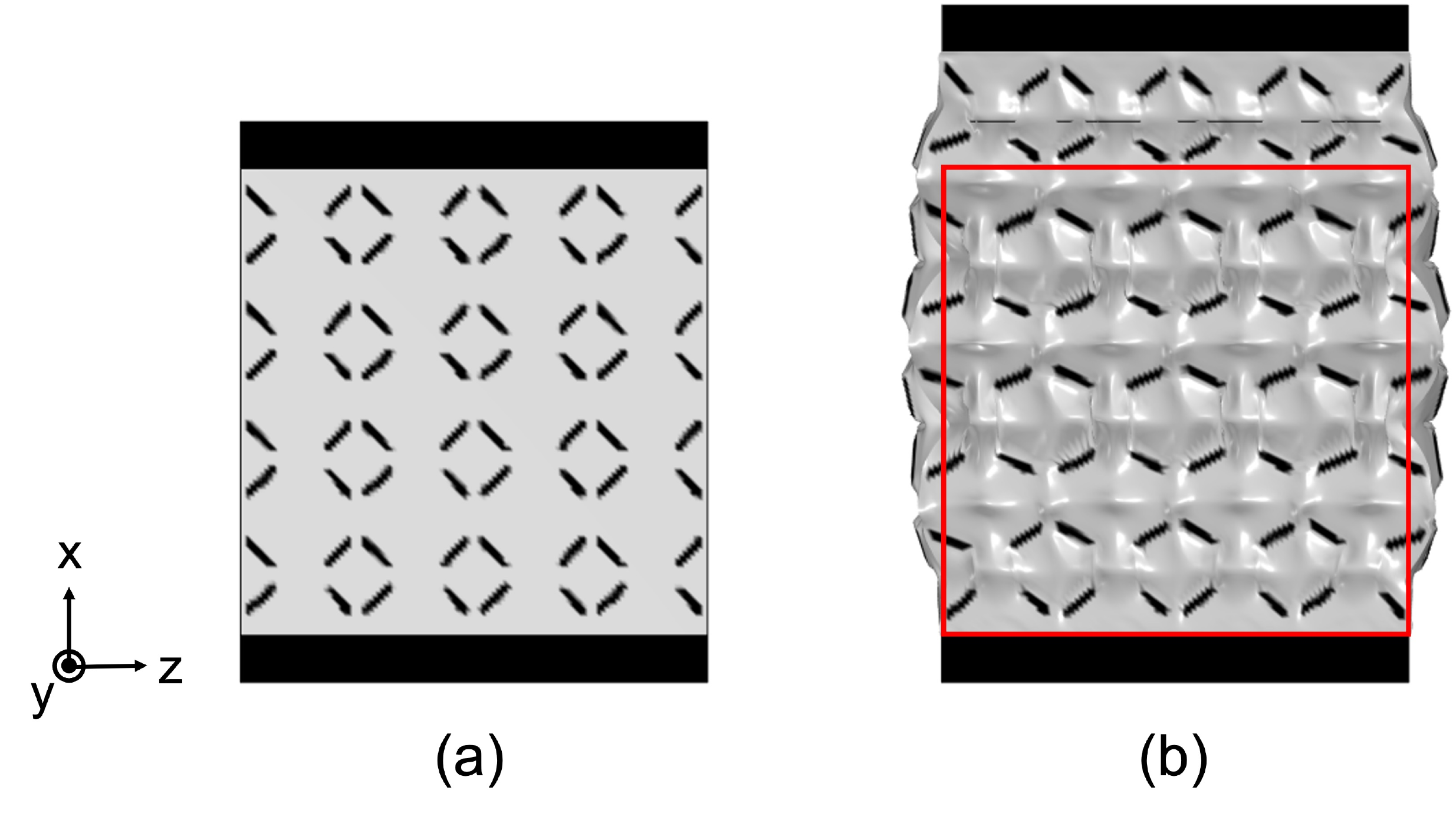}
		\caption{Validation of a 3D metamaterial with negative PR without cavity region.}
		\label{fig:3D-novoid-comsol}
	\end{center}
\end{figure}

\begin{figure}[H]
	\begin{center}
		\includegraphics[scale=0.45]{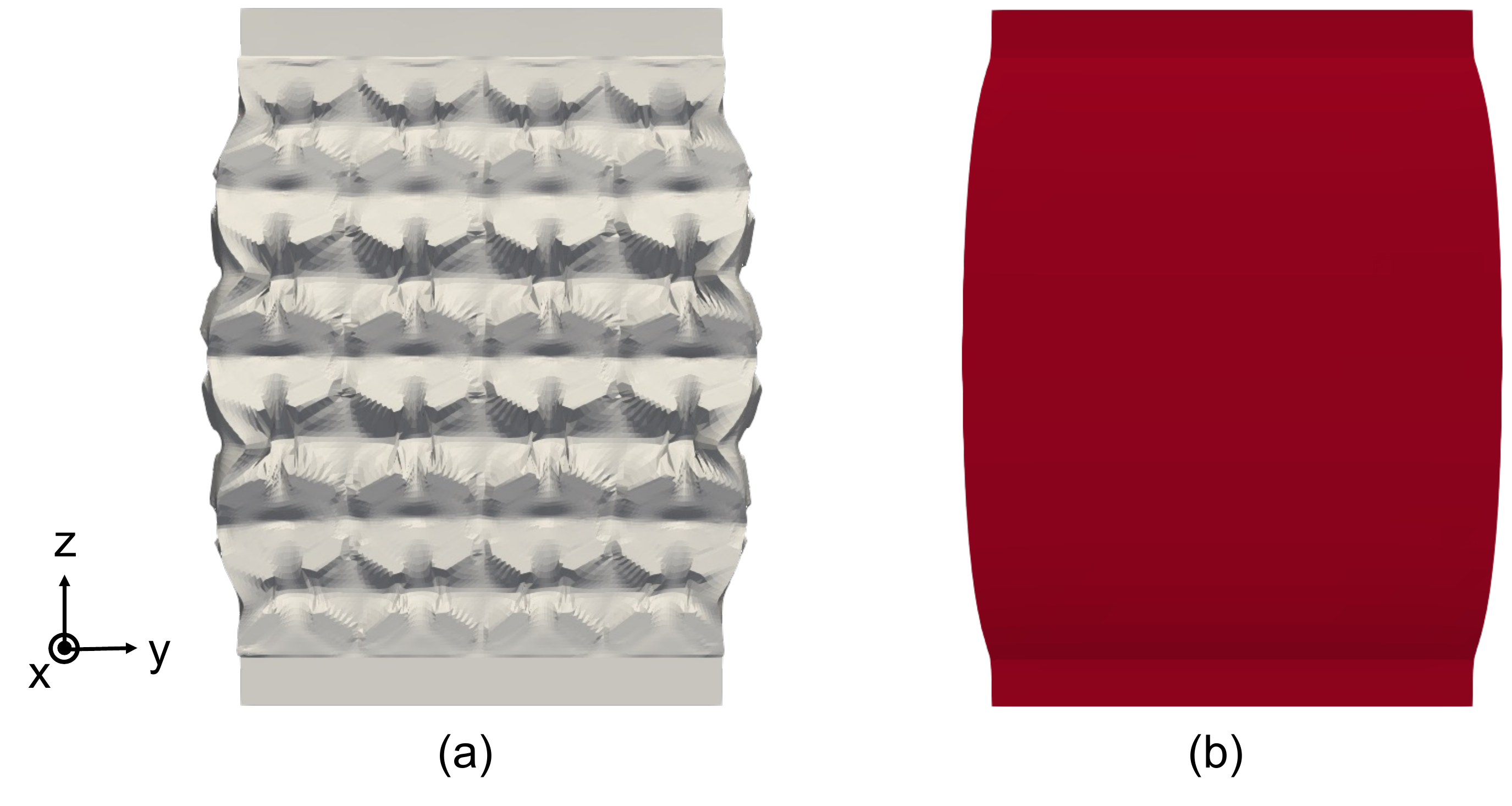}
		\caption{
			Validation of deformation behavior of 3D metamaterials with negative PR when subjected to tensile displacement; (a) structure comprising homogeneous material and (b) multiple array of unit cells.
		}
		\label{fig:coef_valid}
	\end{center}
\end{figure}

Fig. \ref{fig:3D-novoid-comsol}(b) shows the behavior when tensile boundary displacements along the $x$-axis are applied to the arrays of the optimal structure shown in Fig. \ref{fig:3D-novoid-comsol}(a) using the same method as in the previous subsection.
As Subsection 5.3.1, a negative PR is realized.
Fig.~\ref{fig:3D-novoid-comsol} shows that when strain is applied in the tensile direction along the $x$-axis, orthogonal direction strain occurs because of the opening of the structure of material 1 along the $z$-axis, and the entire structure expands. 
For the transverse strain, the deformation in this subsection is larger than that of the previous subsection.
This seems reasonable because the absolute value of PR for the optimal structure with cavities was approximately $0.3$, whereas that of the optimal structure without cavities was approximately $0.4$.
In addition, the validity of the homogenized coefficients was verified by comparing the deformed shape in Fig.~\ref{fig:3D-novoid-comsol}(b) with that of a homogeneous material 
with the homogenized elasticity tensor obtained by this optimization when subjected to similar tensile displacements.
The white structure in Fig.~\ref{fig:coef_valid}(a) represents the shape shown in Fig.~\ref{fig:3D-novoid-comsol}(b), and the red structure in Fig.~\ref{fig:coef_valid}(b) represents the shape of the homogeneous material.
Both of them show expansion in the y-direction in response to tensile strain in the x-direction, and indeed, the deformed shapes are generally similar. 
It is considered that increasing the number of unit cell arrays will reduce the undulation of the external shape in Fig.~\ref{fig:coef_valid}(a) and bring it closer to the shape in Fig.~\ref{fig:coef_valid}(b).
The above results indicate the validity of the homogenized coefficients.

\section{Experimental validation}
The 3D structures without cavities obtained from optimization are stacked using a 3D printer. 
We conducted tensile experiment to confirm the validity of the optimization.

In this study, we used Objet260 Connex3 from Stratasys as a 3D printer. 
As mentioned above, VeroClear was used as material 1 with high stiffness and TangoPlus as material 2 with low stiffness. 
Both materials are resin materials marketed by Stratasys.

\begin{figure}[H]
	\begin{tabular}{cc}
		\begin{minipage}[t]{0.45\hsize}
			\centering
			\includegraphics[scale=0.4]{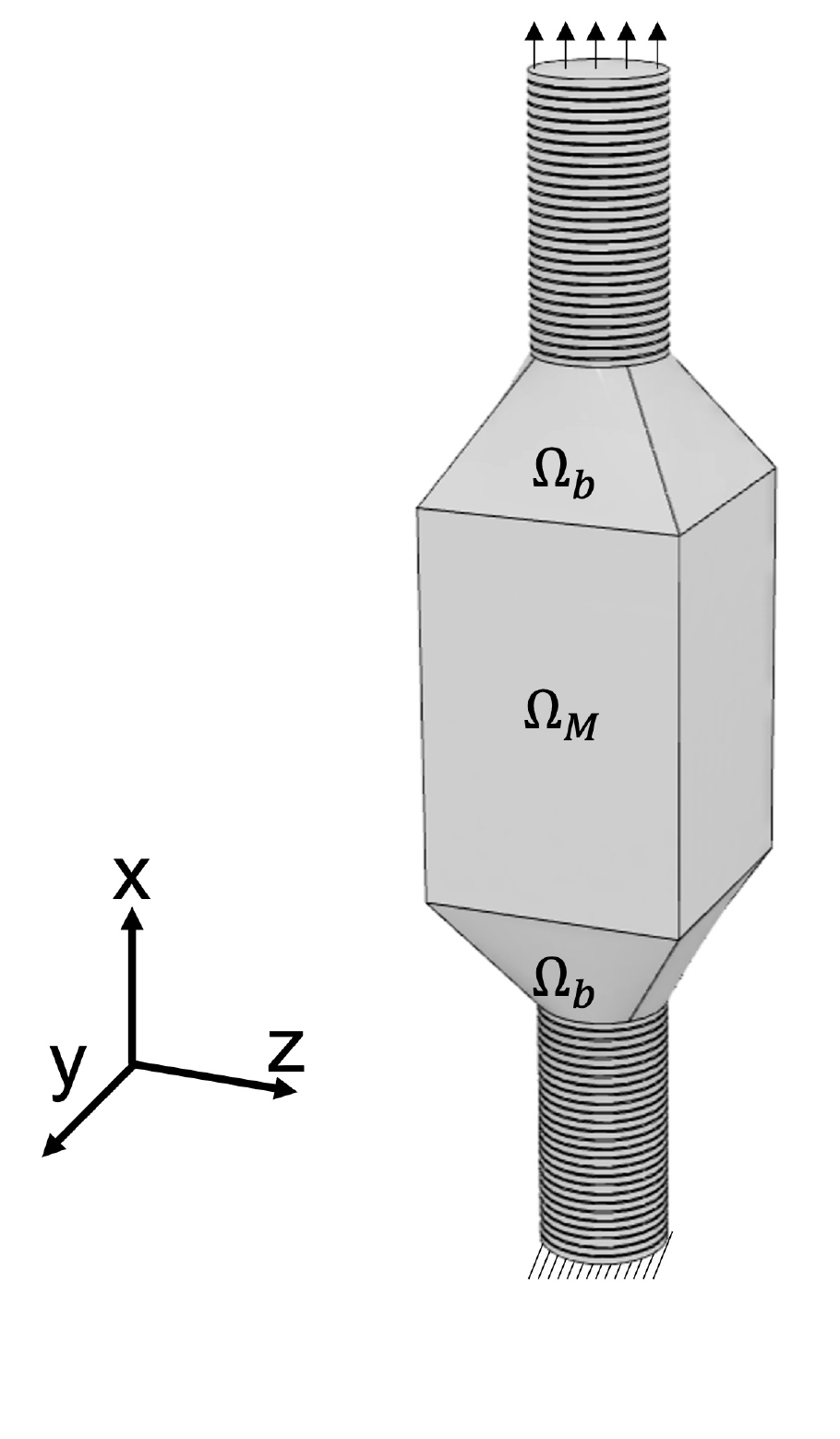}
			\caption{Model for validating optimal structure. }
			\label{fig:model-ex}
		\end{minipage}&
		\begin{minipage}[t]{0.45\hsize}
			\centering
			\includegraphics[scale=0.4]{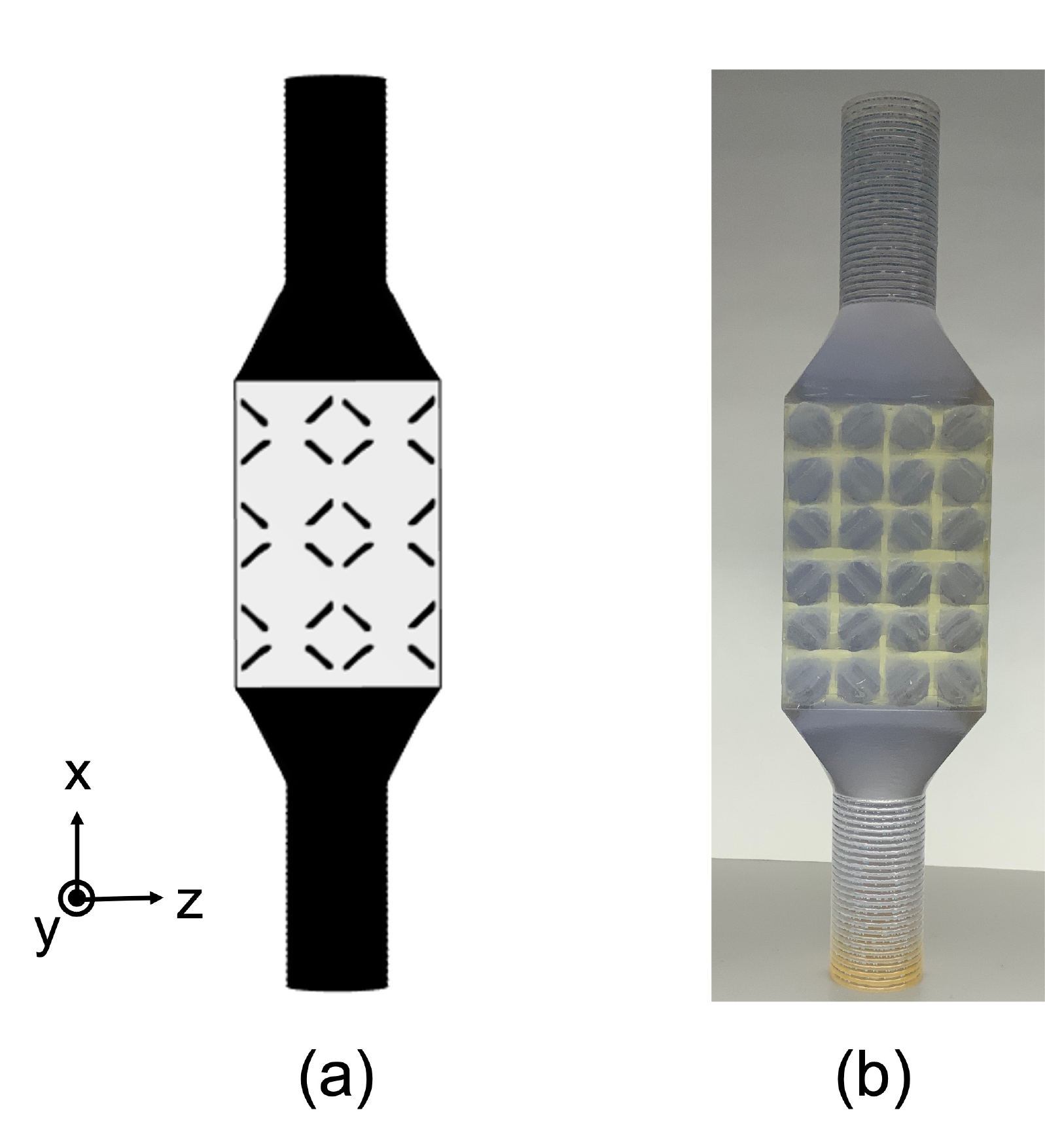}
			\caption{(a)The 3D model established using CAD software and (b)the 3D-printed sample.}
			\label{fig:sample}
		\end{minipage}
	\end{tabular}
\end{figure}

A sample of the 3D structure with negative PR was produced by a 3D printer using the abovementioned materials.
The shape obtained from the proposed method could be manufactured without any modification. 
The sample size was 40 $\times$ 40 $\times$ 120 $mm^3$; the sample comprises $\Omega_M$ and $\Omega_b$ (Fig.~\ref{fig:model-ex}).
The size of the domain $\Omega_M$, which is the domain where the unit cells are arranged, is 40 $\times$ 40 $\times$ 60 $mm^3$; within this domain, the unit cells are periodically arranged in 2, 2,  and 3 cells in the $x$-, $y$-, and $z$-axes, respectively.
The region $\Omega_b$ is a cylindrical region comprising VeroClear and grooved for the chuck of the tensile testing machine.
Fig.~\ref{fig:sample}(a) shows the 3D model established using computer-aided design (CAD) software, and Fig.~\ref{fig:sample}(b) shows the 3D-printed sample.

\begin{figure}[H]
	\begin{center}
		\includegraphics[scale=0.05]{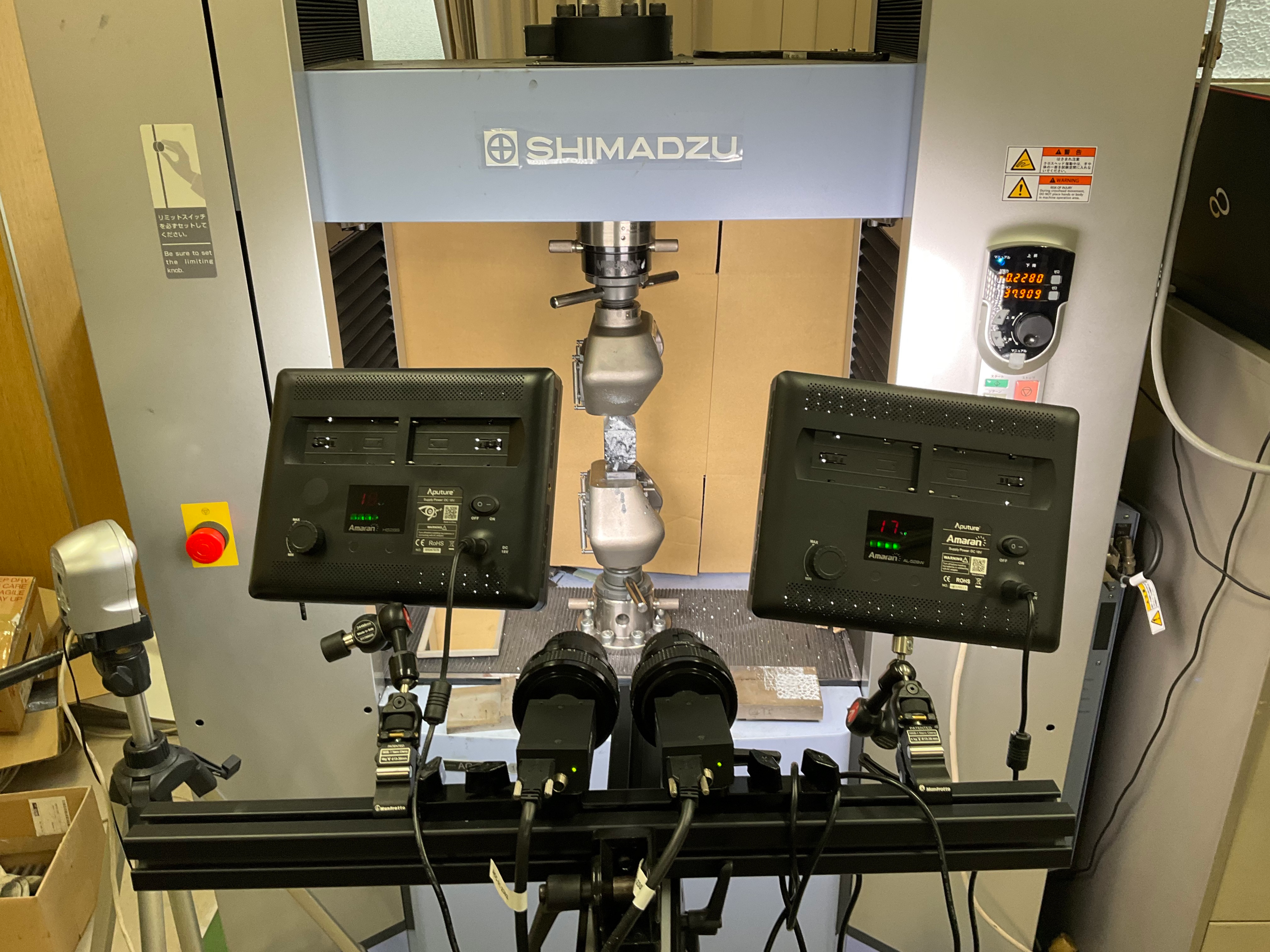}
		\caption{Tensile testing machine and the scene of tensile testing.}
		\label{fig:test-scene}
	\end{center}
\end{figure}

To examine the mechanical properties of the 3D structure with negative PR, we performed the static tension test of the sample with a constant strain rate of 6 mm/min using a 25-tonf screw-driven type universal testing instrument (SHIMADZU, Autograph AG-Xplus) (Fig. \ref{fig:test-scene}). 
We used the digital image correlation (VIC-3D) system to capture the full-field deformation and strain.
The experimental scene is shown in Fig. \ref{fig:test-scene}.
This optimization can be validated from the deformation behavior.

\begin{figure}[H]
	\begin{center}
		\includegraphics[scale=0.7]{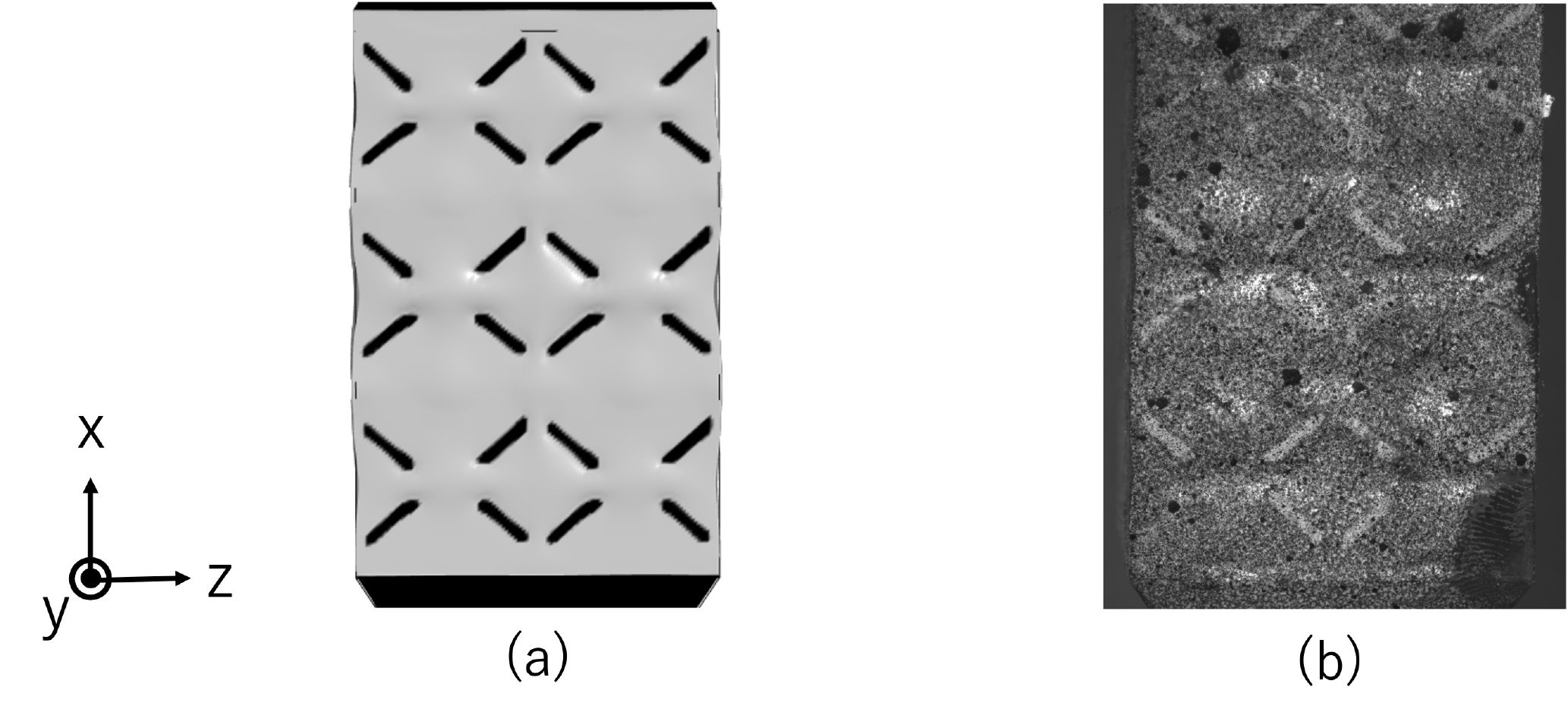}
		\caption{Deformation diagram when a tensile load is applied: (a) deformation from numerical simulation and (b) deformation during the actual test. }
		\label{fig:experiment}
	\end{center}
\end{figure}

\begin{figure}[H]
	\begin{center}
		\includegraphics[scale=0.7]{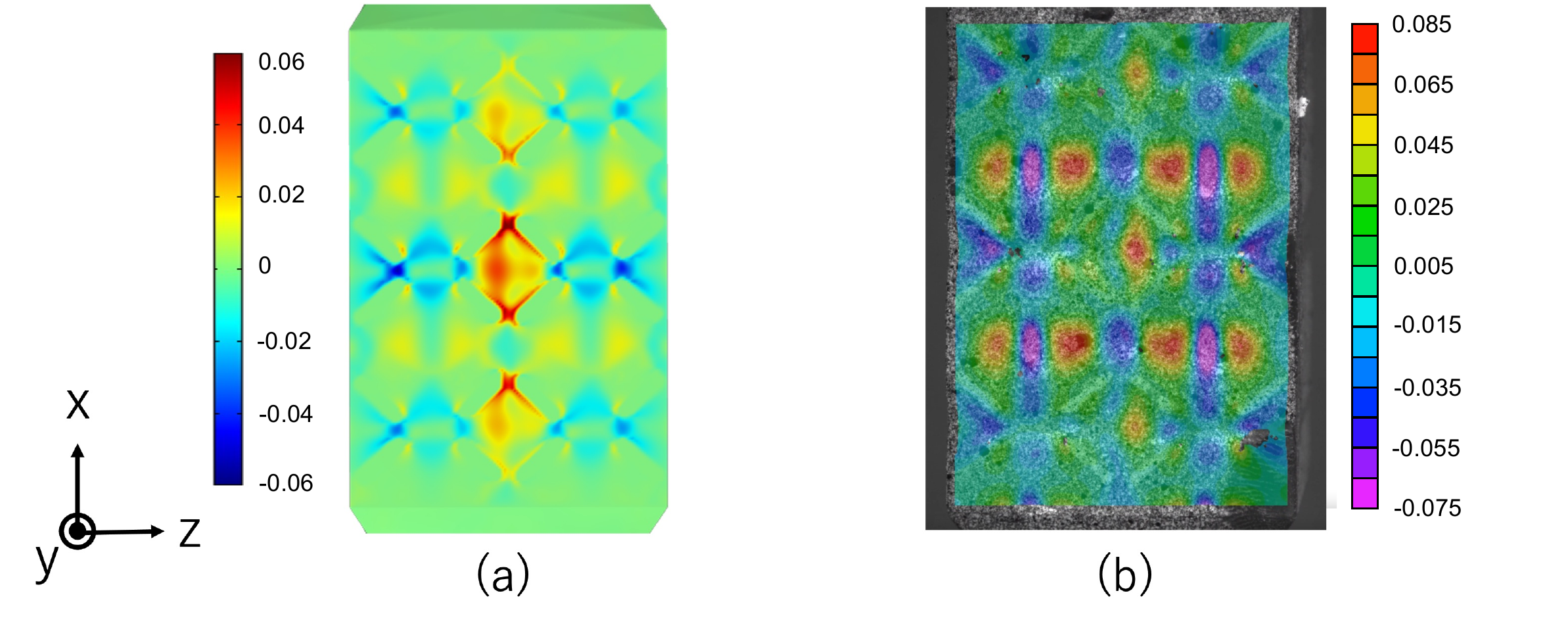}
		\caption{Contour plot of z-axis strain on the model surface: (a) numerical analysis using FEM calculations and (b) analysis using VIC-3D. }
		\label{fig:contour}
	\end{center}
\end{figure}

Fig.~\ref{fig:experiment}(a) and (b) show the deformation of the sample when boundary tensile loads of 6.0 kN and 5.5 kN are applied to the model using FEM calculations and tensile testing, respectively. 
Fig.~\ref{fig:experiment}(a) shows that the neck and expansion shapes alternately appear on the left and right sides, which can be seen in Fig. \ref{fig:experiment}(b); this indicates that the deformation modes are similar.

Fig.~\ref{fig:contour}(a) shows a contour plot of the z-axis vertical strain $e_{zz}$ of the model surface analyzed using FEM calculations. 
Fig.~\ref{fig:contour}(b) shows the contour plot of the $z$-axis vertical strain of the sample, i.e., the vertical strain in the paper direction, measured by VIC-3D. 

Figs.~\ref{fig:contour}(a) and (b) show that the magnitudes of the absolute values of $e_{zz}$ are also similar.
Figs.~\ref{fig:contour}(a) and (b) show that the strain distribution is negative and positive in the materials 1 and 2, respectively, thus demonstrating a similar strain distribution. 
This indicates that material 1, with its higher stiffness, exhibits normal positive PR behavior, whereas material 2, with its lower stiffness, exhibits negative PR behavior, thus resulting in a negative PR behavior of the entire structure.

In summary, the deformation behaviors and the strain distributions are similar for the calculated and experimental results, demonstrating the validity of the optimal structure of the two-phase material and cavity-free metamaterials obtained in this study.
The proposed method was validated by verifying the agreement between the numerical analysis and the experimental results, although $\varepsilon$ is treated as a finite value.

\newpage
\section{Conclusion}
In this paper, we propose a topology optimization method based on the homogenization and level set methods to realize a 3D structure without cavities showing negative PR, which can be fabricated using a 3D printer. 
By applying the optimization consisting of two steps, the proposed optimization method can obtain a three-dimensional structure that does not contain cavities exhibiting negative Poisson's ratio.
We demonstrated the validity of the proposed method by analyzing the behavior of metamaterials based on the optimal structures when subjected to longitudinal strain.
Based on the obtained optimal structures, additive manufacturing and tensile strain tests were conducted to confirm this optimization.
Both numerical analysis and experimental results confirmed the optimization.
In future studies, to enhance performance, we will use multi-material topology optimization, as described in \cite{Noda}, to create cavity-free negative PR metamaterials with three or more phases.
\section*{Acknowledgments}
This work was partly supported by Katsu start-up fund of the University of Tokyo.

The authors would like to thank Enago (www.enago.jp) for the English language review.

\section*{Data availability statement}
Data on the shapes used in the validation of the experiment are available on Mendeley Data (DOI: 10.17632/96w9wxp42w.1).
The other data will be made available on request.


\end{document}